\documentclass[%
 reprint,
superscriptaddress,
nofootinbib,
 amsmath,amssymb,
 aps,
]{revtex4-1}

\usepackage{enumitem}
\usepackage{graphicx}% Include figure files
\usepackage{dcolumn}% Align table columns on decimal point
\usepackage{bm}% bold math
\usepackage{tensor}
\usepackage{hyperref}
\usepackage{color}
\usepackage{soul}
\setstcolor{red}
\usepackage[normalem]{ulem}
\def\ba{\begin{eqnarray}}
\def\ea{\end{eqnarray}}
\newcommand{\mpl}{M_{\rm{Pl}}}
\newcommand{\tpl}{t_{\rm{Pl}}}

\newcommand{\K}{{\cal K}}

\newcommand{\M}{{\cal M}}

\begin{document}

\preprint{}

\title{Instability of spherically-symmetric black holes in Quadratic Gravity}% Force line breaks with \\
%\thanks{A footnote to the article title}%

\author{Aaron Held}
\email{aaron.held@uni-jena.de}
\affiliation{
Theoretisch-Physikalisches Institut, Friedrich-Schiller-Universit\"at Jena,
Max-Wien-Platz 1, 07743 Jena, Germany
}
\affiliation{
The Princeton Gravity Initiative, Jadwin Hall, Princeton University,
Princeton, New Jersey 08544, U.S.
}
\affiliation{
Theoretical Physics, Blackett Laboratory, Imperial College London,
SW7 2AZ London, U.K.
}
\author{Jun Zhang}%
\email{zhangjun@ucas.ac.cn}
\affiliation{International Centre for Theoretical Physics Asia-Pacific, University of Chinese Academy of Sciences, 100190 Beijing, China}
\affiliation{Taiji Laboratory for Gravitational Wave Universe, University of Chinese Academy of Sciences, 100049 Beijing, China}
\affiliation{
Theoretical Physics, Blackett Laboratory, Imperial College London,
SW7 2AZ London, U.K.
}

%\date{\today}% It is always \today, today,
             %  but any date may be explicitly specified

\begin{abstract}
We investigate the linear stability of the two known branches of spherically-symmetric black holes in Quadratic Gravity. We extend previous work on the long-wavelength (Gregory-Laflamme) instability of the Schwarzschild branch to a corresponding long-wavelength instability in the non-Schwarzschild branch. In both cases, the instability sets in below a critical horizon radius at which the two black-hole branches intersect. This suggests that classical perturbations enforce a lower bound on the horizon radius of spherically-symmetric black holes in Quadratic Gravity.
\end{abstract}

%\keywords{Suggested keywords}%Use showkeys class option if keyword
                              %display desired
\maketitle

%\tableofcontents

%======================================
\section{Introduction}
\label{sec:intro}
%======================================

The 1960s and 70s have been referred to as the `golden age' of General Relativity (GR) and have led to a solid theoretical understanding and mainstream acceptance of black holes as astrophysical objects.
Now we are entering another `golden age' in which we gain direct observational access to said black holes. Gravitational-wave interferometers~\cite{LIGOScientific:2016aoc, LIGOScientific:2020ibl} and very long baseline interferometry%(VLBI)
~\cite{EventHorizonTelescope:2019dse, EventHorizonTelescope:2022xnr} respectively observe gravitational-wave and electromagnetic-wave signals which originate from close to the horizon.
This provides a novel opportunity to test whether the astrophysical black holes that we observe agree with the predictions of GR.\\

Ideally, we want to test the predictions of GR against the predictions of theories beyond GR. These theories are motivated by the cosmological riddles of dark matter~\cite{Bertone:2004pz, Barack:2018yly} and dark energy~\cite{Peebles:2002gy}, from the breakdown of classical GR in the black-hole interior~\cite{Penrose:1964wq, Wald:1997wa}, as well as by quantum fluctuations, see e.g.~\cite{Donoghue:1994dn}.
Focusing on metric theories governed by local actions, we can broadly classify modifications of GR into those that involve additional matter fields and those stemming from higher-order curvature operators (see, for example, Ref.~\cite{Clifton:2011jh} for a review). In this work, we focus on the latter and, in particular, on modifications quadratic in curvature. 
\\

Quadratic-curvature operators are generally expected in gravitational theories beyond GR and are motivated from two different points of view.
From the effective field theory (EFT) point of view, the quadratic terms serve as the leading-order corrections to the Einstein-Hilbert term in the EFT expansion of an infinite tower of higher-dimensional operators. Such higher-dimensional operators, including the quadratic ones, capture effects from potential UV physics, e.g., all the unknown fields with masses above the EFT cut-off scale. The EFT is valid as long as the higher-dimensional operators are suppressed by powers of the cut-off scale, and hence contribute only perturbatively. 
Contrary to the EFT point of view, quadratic curvature operators can also arise as a fundamental modification of GR -- possibly motivated by quantum gravity~\cite{Stelle:1976gc,Avramidi:1985ki,Boulware:1985wk, Zwiebach:1985uq,vandeVen:1991gw,Alvarez-Gaume:2015rwa} -- leading to the so-called theory of Quadratic Gravity~\cite{Stelle:1977ry}. \\

It will be important to distinguish these two points of view: As we will review, the presence of the quadratic operators leads to additional massive degrees of freedom~\cite{Stelle:1977ry,Hindawi:1995an,Hinterbichler:2015soa}: In addition to the massless spin-2 degree of freedom in GR, the quadratic-curvature operators generally propagate a massive spin-0 and a massive spin-2 degree of freedom. These additional degrees of freedom can play an important role in the stability of black holes in Quadratic Gravity, i.e., in the fundamental interpretation, but cannot be excited within the validity of the EFT since their masses are comparable to or larger than the EFT cutoff scale. 
Similarly, (some of) the alternative background solutions (see below) may only occur beyond the regime of validity of the EFT.
% Aaron: It was 'The same distinction applies to the discussion of alternative background solutions (see below) that have no counterpart in GR and the EFT.' I changed this because we are not sure that the non-Schwarzschild branch far above the branch point is not potentially within the validity of the EFT.
Here, we focus mostly on Quadratic Gravity as a fundamental theory and comment on the EFT interpretation.\\

\begin{figure}
    \centering
    \includegraphics[width=\linewidth]{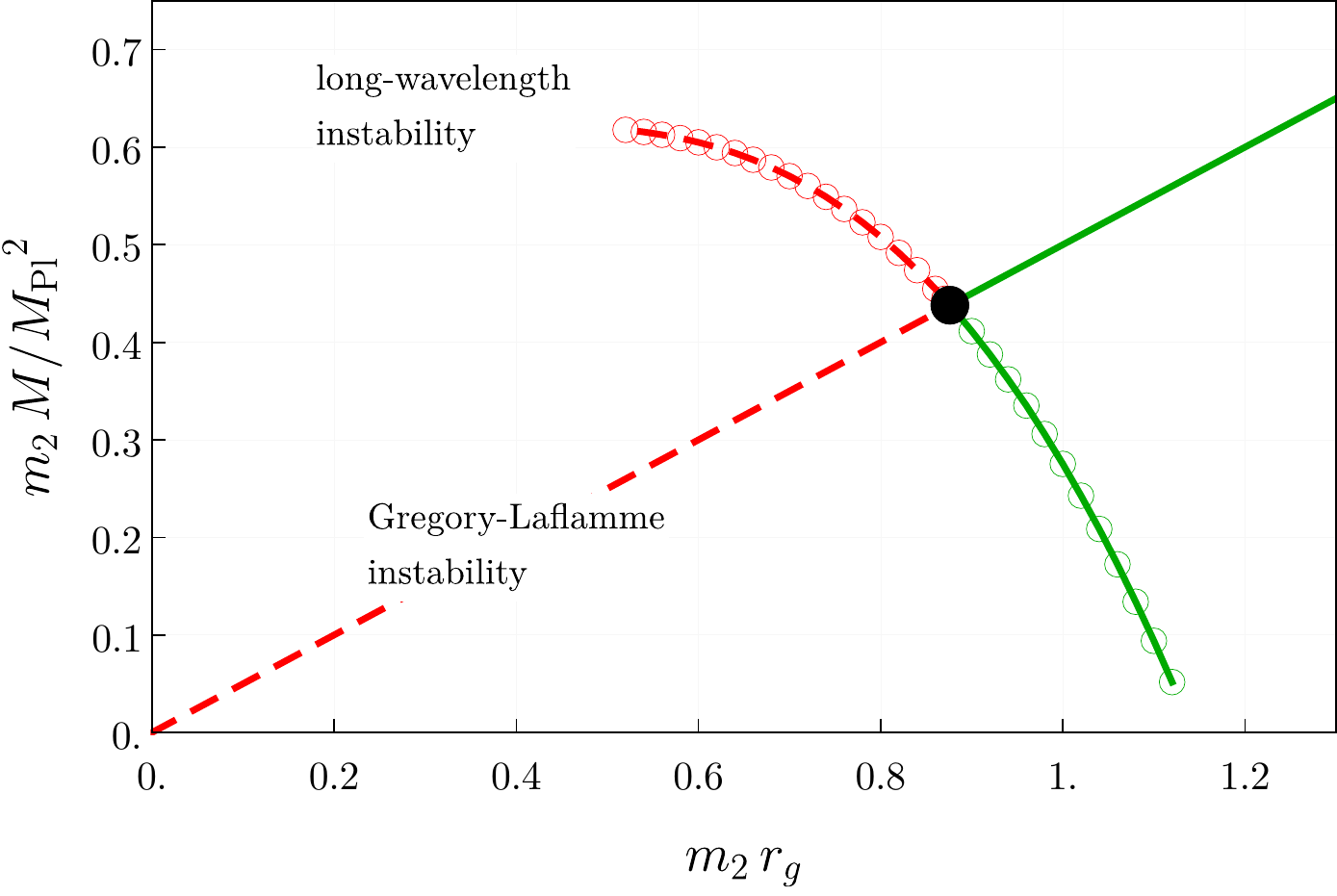}
    \caption{Parametric plot of the Schwarzschild (without circles) and the non-Schwarzschild (with circles indicating numerical data points) black-hole branch as a function of horizon radius $r_g$ and asymptotic mass $M$. The plot axes are re-scaled with $m_2$ (and appropriate powers of $\mpl$) such that the plot applies to any $m_2$.
To the left of the branch point (filled black circle) both black holes are unstable due to a classical long-wavelength instability (cf.~Sec.~\ref{sec:stability}). To the right of the branch point, Schwarzschild spacetime is unstable due to Hawking radiation. If the same holds for the non-Schwarzschild branch, as suggested by thermodynamic arguments (see~\cite{Fan:2014ala, Goldstein:2017rxn, Lu:2017kzi}), the branch point may be an attractor, see discussion in Sec.~\ref{sec:dicussion}. 
}
    \label{fig:syopsis}
\end{figure}

According to Birkhoff's theorem, the static and spherically symmetric vacuum solution of GR is uniquely described by the Schwarzschild metric. While the Schwarzschild solution is also a vacuum solution of Quadratic Gravity, additional branches of vacuum solutions, both black holes and horizonless objects, have been found~\cite{Holdom:2002xy,Daas:2022iid,Lu:2015psa,Lu:2015cqa,Lu:2015tle,Kokkotas:2017zwt,Pravda:2016fue,Lu:2017kzi,Podolsky:2019gro,Daas:2022iid}. Quadratic Gravity, therefore, breaks the uniqueness theorem of vacuum GR. %
In particular, Quadratic Gravity admits a second branch of static, spherically symmetric, and asymptotically flat black-hole spacetimes in addition to the Schwarzschild branch. These two spherically symmetric black-hole branches can be represented in terms of their horizon radius $r_g$ and their asymptotic mass $M$. While the Schwarzschild branch is represented by $r_g=2GM$ (cf.~line in Fig.~\ref{fig:syopsis}) and is Ricci-flat, the other branch (cf.~open circles in Fig.~\ref{fig:syopsis}) has no known closed analytical form and is no longer Ricci-flat (with the exception of the branch point). The two branches intersect at a branch point at which $2GMm_2 = r_g m_2 \equiv p \approx0.87$, where $m_2$ is the mass of the massive spin-2 degree of freedom and is determined by the coefficients of the quadratic operators.\\

Before exploring the phenomenology of such alternative black-hole branches, it is pertinent to understand their stability. As for classical stability, it is known that a linear long-wavelength instability has been found for small-mass black holes in the Schwarzschild branch~\cite{Brito:2013wya,Lu:2017kzi}. Specifically, the Schwarzschild black holes with masses below the branch point ($M<p/2Gm_2$ or $r_g < p/m_2$) are unstable against spherically symmetric perturbations, which are linearly equivalent to the well known Gregory-Laflamme instability~\cite{Gregory:1993vy,Gregory:2011kh,Collingbourne:2020sfy} of higher-dimensional black strings. 

In this paper, we review that this linear long-wavelength instability of the Schwarzschild black hole is associated with the monopole perturbations of the massive spin-2 degree of freedom of Quadratic Gravity that is manifest in the Einstein frame. We then extend the linear-stability analysis to the non-Schwarzschild branch and find a similar long-wavelength instability that occurs whenever the horizon radius of the non-Schwarzschild black hole exceeds the horizon radius of the black hole at the branch point, i.e., whenever $r_g < p/m_2$ or equivalently $M>p/2Gm_2$. Note that, in contrast to the Schwarzschild branch, this means that the instability occurs for black holes with large asymptotic mass.
\\

The rest of this paper is structured as follows: In Sec.~\ref{sec:QG}, we review the field equations~(Sec.~\ref{sec:eoms}), the degrees of freedom~(Sec.~\ref{sec:dofs}), and the static spherically-symmetric black-hole solutions~(Sec.~\ref{sec:bg}) of Quadratic Gravity. 
In Sec.~\ref{sec:perts}, we linearize the dynamics~(Sec.~\ref{sec:RicciScalarFlat} and~\ref{sec:RicciFlat}), decompose into spherical harmonics~(Sec.~\ref{sec:monopole-perts}), and derive the key technical result of our work: the master equation for monopole perturbations on arbitrary static and spherically-symmetric backgrounds.
In Sec.~\ref{sec:stability}, we recover the well-known Gregory-Laflamme instability~\cite{Gregory:1993vy,Gregory:2011kh,Collingbourne:2020sfy} of Schwarzschild black holes~(Sec.~\ref{sec:GF-instab}) and find a similar long-wavelength instability in the non-Schwarzschild branch~(Sec.~\ref{sec:new-instab}). We end with a discussion in Sec.~\ref{sec:dicussion} and delegate several technical complications to an appendix. We work in mostly-plus signature and in units in which $\hbar = c = 1$. 

%
%======================================
\section{The theory of Quadratic Gravity}
\label{sec:QG}
%======================================

There are three possible operators that are quadratic in Riemann curvature, namely $R_{\mu\nu\rho\sigma}R^{\mu\nu\rho\sigma}$, $R_{\mu\nu}R^{\mu\nu}$, and $R^2$, where $R_{\mu\nu\rho\sigma}$, $R_{\mu\nu}$ and $R$ denote the Riemann tensor, the Ricci tensor and the Ricci scalar respectively. In four dimensions, the Gauss-Bonnet invariant $\mathcal{G} = R_{\mu\nu\rho\sigma}R^{\mu\nu\rho\sigma} - 4 R_{\mu\nu}R^{\mu\nu} + R^2$ is a total derivative and thus topological. This leaves two independent operators at quadratic order in curvature. Without loss of generality, the action of (vacuum) Quadratic Gravity can be written as
\begin{align}
\label{eq:QG-action}
S = \mpl^2 \int d^4 x \sqrt{-g}\Bigg[
	&\frac{1}{2}R
	+\frac{1}{12m_0^2}R^2
	\\\notag &
	-\frac{1}{4m_2^2}C_{\mu\nu\alpha\beta}C^{\mu\nu\alpha\beta}
\Bigg]
+S_\text{matter}
\;,
\end{align}
where $\mpl$ is the Planck mass, $C_{\mu\nu\alpha\beta}$ is the Weyl tensor, and we use the shorthand notation $\sqrt{-g} \equiv \sqrt{-\det{g}}$. As we will see below, the parameters $m_0$ and $m_2$ are associated with the masses of a spin-0 and a spin-2 degree of freedom (in addition to the massless spin-2 degree of freedom of GR).
\\

%
%--------------------------------------
\subsection{Field equations}
\label{sec:eoms}
%--------------------------------------

The field equations following from the action in \eqref{eq:QG-action} (see e.g.~\cite{Lu:2017kzi}) read 
\begin{align}
\label{eq:eom}
    \frac{T_{\mu\nu}}{\mpl^2}=\;
    &G_{\mu\nu}
    -\frac{1}{3}\left(
        \frac{1}{m_0^2} 
        - \frac{1}{m_2^2}
    \right)\left(
        D_\mu D_\nu
        -g_{\mu\nu}\Box
    \right)R
    \notag\\&
    -\frac{1}{m_2^2}\,\Box\,G_{\mu\nu}
    -\frac{2}{m_2^2}\left(
        R_{\mu \rho \nu \sigma}
        -\frac{1}{4}g_{\mu\nu}\,R_{\rho \sigma}
    \right)R^{\rho \sigma}
    \notag\\&
    +\frac{1}{3}\left(
        \frac{1}{m_0^2} 
        + \frac{2}{m_2^2}
    \right)\left(
        R_{\mu\nu}
        -\frac{1}{4}g_{\mu\nu}R
    \right)R
    \;,
\end{align}
where $T_{\mu\nu}$ is the energy-momentum tensor associated to $S_\text{matter}$ in~\eqref{eq:QG-action}.

In the present work, we specify to vacuum solutions with $T_{\mu\nu}=0$. In this case, taking the trace of \eqref{eq:eom} gives
\begin{align}
    \frac{1}{m_0^2}\,\Box R - R = 0\;.
\end{align}
This relation has been used to prove that $R=0$ must hold for \emph{any} static, spherically-symmetric, and asymptotically-flat vacuum solution of Quadratic Gravity~\cite{Nelson:2010ig, Lu:2015cqa}.
Put differently, static, spherically-symmetric, and asymptotically-flat vacuum solutions of Quadratic Gravity are ``Ricci-scalar flat'': A special subclass of the former are the Ricci-flat ($R_{\mu\nu}=0$) vacuum solutions of GR.
\\

%--------------------------------------
\subsection{Dynamical degrees of freedom}
\label{sec:dofs}
%--------------------------------------

The field equations \eqref{eq:eom} are fourth order in derivatives.
However, they can be reduced to second order by introducing auxiliary fields. We follow Ref.~\cite{Hinterbichler:2015soa} (see also Ref.~\cite{Hindawi:1995an}) and introduce the auxiliary fields at the level of the action.

First, we can remove the $R^2$ term by introducing a scalar field $\phi$ and write \eqref{eq:QG-action} as
\begin{align}
\label{eq:QG-action-1}
    S = \mpl^2 \int d^4 x \sqrt{-g}
    &\Bigg[ 
        \frac{1}{2}\left(1+\frac{\phi}{3m_0^2}\right) R 
        - \frac{1}{12m_0^2} \phi^2
        \notag\\&
        -\frac{1}{4m_2} C_{\mu\nu\alpha\beta}C^{\mu\nu\alpha\beta}
    \Bigg].
\end{align}
Varying \eqref{eq:QG-action-1} with respect to $\phi$ yields the equation of motion $\phi=R$ for the auxiliary field, by which one recovers~\eqref{eq:QG-action}. 
Second, we can perform a conformal transformation, which (by definition) leaves the Weyl-squared term invariant, i.e.,
\ba
g_{\mu\nu} \rightarrow \frac{3\,m_0^2}{\phi+3\,m_0^2}\, g_{\mu\nu}\, .
\ea
In addition, we redefine $\phi = 3 m_0^2 \left(e^{\psi}-1\right)$, so that the scalar field $\psi$ takes a canonical form in the Einstein frame, i.e.,
\begin{align}
    \label{eq:QG-action-2}
    S = 
    \mpl^2 \int d^4 x \sqrt{-g}&\Bigg[ 
        \frac{1}{2} R 
        %-\frac{3}{4} {\pd \psi}^2 
        +\frac{3}{4} \psi\Box\psi 
        - \frac{3}{4}m_0^2 e^{-2\psi}\left(e^{\psi}-1\right)^2
        \notag\\&
        -\frac{1}{4m_2} C_{\mu\nu\alpha\beta}C^{\mu\nu\alpha\beta}
    \Bigg].
\end{align}
In the Einstein frame, the scalar $\psi$ appears as a minimally coupled matter field with a non-trivial potential.
Finally, we can remove the Weyl-squared term by introducing an auxiliary tensor field $f_{\mu\nu}$, for which \eqref{eq:QG-action-2} can be rewritten as
\begin{align}
    \label{eq:QGactionDOFs}
    S = 
    \mpl^2 \int d^4 x &\sqrt{-g}\Bigg[
        \frac{1}{2} R 
        %-\frac{3}{4} {\pd \psi}^2
        +\frac{3}{4} \psi\Box\psi 
        - \frac{3}{4}m_0^2 e^{-2\psi}\left(e^{\psi}-1\right)^2 
        \notag\\&
        + f_{\mu\nu}G^{\mu\nu}
        +\frac{1}{2}m_2^2\left(f_{\mu\nu}f^{\mu\nu}-f^2\right)
    \Bigg].
\end{align}
Again, \eqref{eq:QGactionDOFs} and \eqref{eq:QG-action-2} are equivalent upon using the equation of motion for the auxiliary field $f_{\mu\nu}$ (obtained by varying~\eqref{eq:QGactionDOFs} with respect to $f^{\mu\nu}$ -- see below).
Thereby, we have recast the theory into manifestly second-order form. The decoupling limit confirms that the theory contains a massless spin-2 field $g_{\mu\nu}$ (as does GR), a spin-0 field $\psi$ with mass $m_0$, and a spin-2 field $f_{\mu\nu}$ with mass $m_2$~\cite{Stelle:1977ry,Hinterbichler:2015soa}. The massive spin-2 field comes with an opposite-sign kinetic term (compared to the other fields) and is thus an Ostrogradski ghost.\\

The equations of motion for $\psi$ (obtained by variation w.r.t. $\psi$ itself) and $g_{\mu\nu}$ (obtained by variation w.r.t. $f^{\mu\nu}$) respectively read
\begin{align}
    \label{eq:variation-spin0}
    \Box\psi + m_0^2\,e^{-2\psi}(e^\psi-1) 
    &= 0\;,
    \\[0.5em]
    \label{eq:variation-metric}
    \mathcal{H}_{\mu\nu} 
    \equiv\,
    G_{\mu\nu} 
    + m_2^2 \left(f_{\mu\nu}-f g_{\mu\nu}\right)
    &= 0\;.
\end{align}
As mentioned above, $\mathcal{H}_{\mu\nu}=0$ can be used to re-express the Ricci tensor and Ricci scalar in terms of $f_{\mu\nu}$ and $f$, respectively, i.e.,
\begin{align}
	\label{eq:Ricci-to-f}
	R_{\mu\nu} = -m_2^2 \left(f_{\mu\nu}+\frac{1}{2} f g_{\mu\nu}\right)\;, 
	\quad 
	R=-3\,m_2^2f\;.
\end{align}
Moreover, $D^\nu\,\mathcal{H}_{\mu\nu}=0$ implies (by use of the contracted Bianchi identity $D^\nu\,G_{\mu\nu}=0$) that
\begin{align}
	\label{eq:Bianchi-H-identity}
	D^\nu f_{\mu\nu}=D_\mu f\;.
\end{align}
Using \eqref{eq:Ricci-to-f} and \eqref{eq:Bianchi-H-identity}, as well as partial integration and commutation of covariant derivatives, the equation of motion for $f_{\mu\nu}$ (obtained by variation w.r.t. $g^{\mu\nu}$) can be written as
\begin{align}
    \label{eq:variation-ghost}
    0=\mathcal{F}_{\mu\nu}\equiv\;&
    \Box f_{\mu\nu}
    - D_\mu D_\nu f 
    + 2\,R_{\mu \rho \nu \sigma}f^{\rho \sigma} 
	\\\notag&
	- m_2^2 \left[
		f_{\mu\nu}\left(f -1\right) 
		+ g_{\mu\nu}\left( 
		    f + \frac{1}{2} f_{\rho \sigma}f^{\rho \sigma}
		\right)
	\right]
	\\\notag&
	-\frac{3}{4}\psi\left(
	    \Box - 2\,D_\mu D_\nu
	\right)\psi
	+\frac{3}{4}m_0^2e^{-2\psi}\left(e^{\psi}-1\right)^2
	\;.
\end{align}
The trace of \eqref{eq:variation-ghost} is not a dynamical equation
but rather (after using \eqref{eq:variation-spin0} to remove $\Box\psi$) a constraint, i.e.,
\begin{align}
\label{eq:constraint_f}
    f = \frac{m_0^2}{m_2^2}\,
    e^{-2\psi}\left(e^\psi-1\right)
    \left(e^\psi-1+\frac{1}{2}\psi\right)\;.
\end{align}
Up to here, the equations of motion are fully general. In the special case of $\psi=0$, the constraint \eqref{eq:constraint_f} reduces to $f=0$.
\\

%--------------------------------------
\subsection{Spherically-symmetric black-hole branches}
\label{sec:bg}
%--------------------------------------

In this section, we review the black-hole solutions that are discussed in Refs.~\cite{Lu:2015cqa,Lu:2015psa,Podolsky:2019gro}. Following Ref.~\cite{Lu:2015cqa}, we start with the general metric for static, spherically-symmetric spacetime
\begin{align}\label{eq:metric}
{\rm d}s^2 = - A(r) {\rm d}t^2 + \frac{1}{B(r)} {\rm d}r^2 + r^2 ({\rm d}\theta^2 + \sin^2\theta {\rm d}\varphi^2)\,.
\end{align}
Assuming that the horizon is located at $r_g$, we can expand the metric near the horizon, i.e.,
\ba\label{ABexpansion}
    &&A(r) = a_c\;
    \sum_{n=1}^\infty
    a_n\left(
        \frac{r}{r_g}-1
    \right)^n
    \;,
    \nonumber \\
    &&B(r) = \sum_{n=1}^\infty
    b_n\left(
        \frac{r}{r_g}-1
    \right)^n
    \;,
\ea
where $a_1\equiv1$ such that $a_c$ is a free parameter which be chosen to ensure that $A\rightarrow1$ as $r\rightarrow\infty$. Substituting the ansatz~\eqref{ABexpansion} into the field equations~\eqref{eq:eom}, we can solve for the $a_i$ and $b_i$ with $i \ge 2$ in terms of $b_1$ and $r_g$. For example, for the lowest-order coefficients, we find
\ba
a_2 &=&-\frac{8 \delta ^2-\delta  \left(m^2 r_g^2-12\right)+4}{4 (\delta +1)^2} \,,
\notag\\
b_2 &=&-\frac{8 \delta ^2+3 \delta  \left(m^2 r_g^2+4\right)+4}{4 (\delta +1)}\,,
\ea
where we have defined $b_1 = 1+\delta$. In other words, the black-hole solution is fully determined by $\delta$ and $r_g$. Since the ansatz~\eqref{ABexpansion} is expanded near the horizon, it is not guaranteed that the black hole solution described by $a_i$ and $b_i$ is asymptotically flat for any $\delta$ given a certain $r_g$. When $\delta=0$, we obtain the Schwarzschild solution, which is obviously a solution to the field equation due to its Ricci-flatness. When $\delta \neq 0$, there also exists an asymptomatically flat solution, which we refer to as the non-Schwarzschild black hole. As far as we know, the non-Schwarzschild solution does not have a closed analytical form. In practice, it is obtained numerically, for example by tuning $\delta$ or $r_g$ with the shooting method.

In contrast to Schwarzschild spacetime, the non-Schwarzschild black hole is not Ricci-flat. Moreover, the asymptotic behavior is expected to be
\ba\label{Asymp}
A(r) = 1 - \frac{C_{2,0}}{r} - C_{2-}\frac{e^{-m r}}{r} + \cdots
\ea
where $\cdots$ represents the sub-leading terms \cite{Lu:2015psa}. In order to compare with the Schwarzschild solution, we define an effective ADM mass
\ba\label{ADM}
M \equiv \frac{C_{2,0}}{2G}\,,
\ea
where $C_{2,0}$ can be extracted by fitting $A(r)$ with Eq.~\eqref{Asymp} at large $r$ (e.g., $r\sim 50\, r_g$), once the solution is obtained numerically. Then the two branches of solutions can be represented by plotting $M(r_g)$, see Fig.~\ref{fig:syopsis}. 

While the Schwarzschild horizon increases linearly with the ADM mass, the horizon of the non-Schwarzschild black hole decreases with the ADM mass, and the two branches cross at the branch point with $m_2 r_g = p \approx 0.87$. For $r_g m_2 \gtrsim 1.14$, a distant observer would even observe a negative ADM mass for the non-Schwarzschild black hole. As $r_g$ approaches zero, the Ricci curvature of the non-Schwarzschild solution diverges, leaving a naked singularity (see App~\ref{app:singularity}).

Although the non-Schwarzschild black hole does not have a known closed analytical form, it is still useful to approximate the numerical solution with an analytic expansion. Such an expansion has been studied in Ref.~\cite{Kokkotas:2017zwt}, where the two metric functions $A(r)$ and $B(r)$ are represented by a continued-fraction expansion~\cite{Rezzolla:2014mua}, parameterized by a single dimensionless parameter
\begin{align}
	r_g\,m_2 \lesssim 1.14\;.
\end{align}
The details of the approximation are reviewed in App.~\ref{app:continuedFractionApprox}.
Towards the bound, $r_g\,m_2 \lesssim 1.14$, the mass of the non-Schwarzschild black hole shrinks to zero. 
While~\cite{Kokkotas:2017zwt} restricts to $r_g\,m_2 \gtrsim 0.87$, the non-Schwarzschild solution persists also for smaller values of $r_g\,m_2$ and the continued-fraction expansion remains valid.

%--------------------------------------
\section{Black-hole perturbations}
\label{sec:perts}
%--------------------------------------

We are interested in linear perturbations $\delta\psi$, $\delta g_{ab}$, and $\delta f_{ab}$ about a background $\bar{\psi}$, $\bar{g}_{ab}$, and $\bar{f}_{ab}$, i.e.,
\begin{align}
    \psi &= \bar{\psi} + \delta\psi\;,
    \notag\\
    g_{\mu\nu} &= \bar{g}_{\mu\nu} + \delta g_{\mu\nu}\;,
    \notag\\
    f_{\mu\nu} &= \bar{f}_{\mu\nu} + \delta f_{\mu\nu}\;,
    \label{eq:linear-perturbations}
\end{align}
respectively. Further, we restrict to static, spherically-symmetric black-hole backgrounds.

\subsection{Linear dynamics on ``Ricci-scalar-flat'' backgrounds}
\label{sec:RicciScalarFlat}
We recall that static, spherically-symmetric, and asymptotically flat vacuum solutions of Quadratic Gravity are Ricci-scalar flat, i.e., $\bar{R}=0\Leftrightarrow \bar{f}=0 \Leftrightarrow \bar{\psi}=0$ holds~\cite{Nelson:2010ig, Lu:2015cqa}. In this case, the linear dynamics simplifies to
\begin{align}
    \label{eq:linear-eom-spin0}
    0=\,&
    \bar{\Box}\,\delta\psi 
    + m_0^2\,\delta\psi 
    \;,
    \\[1em]
    \label{eq:linear-eom-metric}
    0=\,&
    \delta G_{\mu\nu} 
    + m_2^2 \left(
        \delta f_{\mu\nu}
        -\bar{g}_{\mu\nu}\,\delta f
    \right)
    \;,
    \\[1em]
    0=\,&
    \bar{\Box}\, \delta f_{\mu\nu}
    - \bar{D}_\mu \bar{D}_\nu \,\delta f 
    + 2\,\bar{R}_{\mu \sigma \nu \rho}\,\delta f^{\sigma \rho}
    + 2\,\bar{f}^{\sigma \rho}\,\delta R_{\mu \sigma \nu \rho}
    \notag\\
    &+ m_2^2 \Big[
        - \delta f_{\mu\nu}
		+ \bar{g}_{\mu\nu}\bar{f}^{\sigma \rho}\,\delta f_{\sigma \rho}
		+ \left(
		    \bar{f}_{\mu\nu}
		    + \bar{g}_{\mu\nu}
		\right)\,\delta f
		\notag\\
		&\quad\quad\quad\quad
		+ \frac{1}{2}\,\bar{f}_{\sigma \rho}\bar{f}^{\sigma \rho}\,\delta g_{\mu\nu}
	\Big]
	\;,
	\label{eq:linear-eom-ghost}
\end{align}
where $\delta G_{\mu\nu}$ and $\delta R_{\mu \sigma \nu \rho}$ denote linear perturbations of $G_{\mu\nu}$ and $R_{\mu \sigma \nu \rho}$ with respect to the metric.

Due to $\bar{\psi}=0$,
the linear perturbations $\delta\psi$ decouple and are governed by a massive scalar wave equation on the respective background. These scalar perturbations can therefore be determined separately. In particular, the massive scalar mode will not alter any conclusions in the coupled (massless and massive) spin-2 sector. For the stability analysis in Sec.~\ref{sec:stability}, we thus only focus on the spin-2 sector.

\subsection{Linear dynamics on Ricci-flat backgrounds}
\label{sec:RicciFlat}

On Ricci-flat backgrounds, we have $\bar{R}_{\mu\nu} = 0\Leftrightarrow \bar{f}_{\mu\nu} = 0$ and thus $\bar{R}=0\Leftrightarrow \bar{f}=0 \Leftrightarrow \bar{\psi}=0$. Furthermore, in this case, the contraction of \eqref{eq:linear-eom-ghost} implies $\delta f=0$. Hence, \eqref{eq:linear-eom-metric} and \eqref{eq:linear-eom-ghost} reduce to
\begin{align}
    \label{eq:linear-Ricciflat-eom-metric}
    0=\,&
    \delta G_{\mu\nu} 
    + m_2^2 \delta f_{\mu\nu}
    \;,
    \\[0.5em]
    0=\,&
    \bar{\Box}\, \delta f_{\mu\nu}
    + 2\,\bar{R}_{\mu \sigma \nu \rho}\,\delta f^{\sigma \rho}
    - m_2^2\,\delta f_{\mu\nu}
	\;,
	\label{eq:linear-Ricciflat-eom-ghost}
\end{align}
In this special case, also $\delta f_{\mu\nu}$ decouples and evolves independently of $\delta g_{\mu\nu}$. When $\delta f_{\mu\nu} = 0$, the equation of $\delta g_{\mu\nu}$ and hence its spectrum such as quasi-normal frequencies are identical to those in GR. This is a direct consequence of the fact that any GR vacuum solution is also a solution of Quadratic Gravity. When $\delta f_{\mu\nu}$ gets excited, the perturbations also include the spectrum of a massive spin-2 field, cf.~\cite{Brito:2013wya}. As shown in ~\cite{Brito:2013wya,Lu:2017kzi}, the monopole mode of the massive spin-2 field suffers from the Gregory-Laflamme instability~\cite{Gregory:1993vy,Gregory:2011kh}. 
\\

We note that~\eqref{eq:linear-eom-spin0}, \eqref{eq:linear-Ricciflat-eom-metric}, and~\eqref{eq:linear-Ricciflat-eom-ghost} describe the linear perturbations of Quadratic Gravity on any Ricci-flat background, in particular, including Kerr black holes. We will discuss the potential instabilities of Kerr black holes in Quadratic Gravity and respective observational constraints in a separate publication.
\\

\subsection{Monopole perturbations}
\label{sec:monopole-perts}
We expect that the monopole perturbation of the massless spin-2 field is pure gauge, just as in GR. Indeed, we will show below that the monopole perturbations can be reduced to a single dynamical degree of freedom corresponding to the massive spin-2 mode.

Before we proceed to derive the respective master equation, we algebraically solve the background equations of motion to re-express all higher-order (radial) derivatives of $A(r)$ and $B(r)$ in terms of 1st- and 0th-order derivatives only.
\\

Linear perturbations as in \eqref{eq:linear-perturbations} can be decomposed into spherical harmonics $Y_{\ell m}(\theta,\varphi)$, a time-dependent part $e^{-i\omega t}$, and a radial mode function. 
Due to the spherical symmetry of the background, we can -- without loss of generality -- focus on the axisymmetric (i.e., $m=0$) perturbations.
Moreover, the focus of this work lies on long-wavelength instabilities: These are expected to be excited in the lowest-lying modes of the decomposed spectrum of spherical harmonics. Hence, we focus on the monopole ($\ell=0 \Rightarrow m=0$) perturbations.
\\

For monopole perturbations, the most general decomposition, cf.~\cite{Regge:1957td, Zerilli:1970se, Motohashi:2011pw}, reduces to
\begin{equation}
\delta g_{\mu\nu} = \begin{pmatrix}
-A H_0 & \hspace{0.2cm}& H_1  & \hspace{0.2cm}& 0 & \hspace{0.2cm}&  0 \\
H_1 & \hspace{0.2cm}& H_2/B  & \hspace{0.2cm}& 0 & \hspace{0.2cm}&0 \\
0 & \hspace{0.2cm}& 0 & \hspace{0.2cm}&r^2 \K  & \hspace{0.2cm}& 0\\
0 & \hspace{0.2cm}& 0 & \hspace{0.2cm}& 0 & \hspace{0.2cm}& r^2 \sin^2\theta \K
\end{pmatrix} e^{-i\omega t} \, ,
\label{eq:evenhab}
\end{equation}
for the massless spin-2 perturbations and to
\begin{equation}
\delta f_{\mu\nu} = \begin{pmatrix}
-A F_0 & \hspace{0.2cm}& F_1  & \hspace{0.2cm}& 0 & \hspace{0.2cm}&  0 \\
F_1 & \hspace{0.2cm}& F_2/B  & \hspace{0.2cm}& 0 & \hspace{0.2cm}&0 \\
0 & \hspace{0.2cm}& 0 & \hspace{0.2cm}&r^2 \M  & \hspace{0.2cm}& 0\\
0 & \hspace{0.2cm}& 0 & \hspace{0.2cm}& 0 & \hspace{0.2cm}& r^2 \sin^2\theta \M
\end{pmatrix} e^{-i\omega t}  \, ,
\label{eq:evenfab}
\end{equation}
for the massive spin-2 perturbations. Herein, $H_{0,1,2}(r)$, $\K(r)$, $F_{0,1,2}(r)$ and $\M(r)$ are eight unknown functions of the radial coordinate $r$. 

The massless and massive spin-2 perturbations, i.e., Eqs.~\eqref{eq:linear-eom-metric} and \eqref{eq:linear-eom-ghost}, correspond to 8 perturbation equations for the above 8 modes. Moreover, $D_\mu f^{\mu\nu}=0$ and $f=0$, correspond to 3 constraints. However, not all of these equations are independent. Instead, as we will see below, they can be reduced to a single Regge-Wheeler type master equation. 
\\

We start by choosing a gauge such that
\begin{align}
	\mathcal{K}=H_{0}=0\;.
\end{align}
Further, we can algebraically solve two of the constraints to express any two of the three modes $F_0$, $F_2$, and $\mathcal{M}$ in terms of the remaining modes. We choose to remove $F_0$ and $F_2$. Finally, we can use the two lowest-order metric-perturbation equations to remove $H_1$ and $H_2$. This leaves us with only two massive spin-2 perturbations, i.e., $F_1$ and $\mathcal{M}$, and one constraint. Once these three equations are fulfilled, all the other equations are automatically fulfilled too. The details of this algebraic reduction are given in App.~\ref{app:algebraicReduction}.

Making use of these relations and defining $\phi(r) = -2\omega\mathcal{M}(r)$ and $\chi(r) = F_1(r)$, we end up with two coupled 2nd-order equations 
\begin{align}
\label{eq:phi}
	\phi ''
	+\phi'\left(\frac{4}{r}+\frac{3 A'}{2 A}-\frac{B'}{2 B}\right)
	+ \frac{\omega^2\,\phi}{A\,B}
	+ V_{\phi\phi}\,\phi
	+ V_{\phi\chi}\,\chi
	&=0\;,
	\\
\label{eq:chi}
	\chi ''
	+\chi'\,\left(\frac{2}{r}+\frac{3 A'}{2 A}+\frac{3 B'}{2 B}\right)
	+ \frac{\omega^2\,\chi}{A\,B}
	+ V_{\chi\chi}\,\chi
	+ V_{\chi\phi}\,\phi
	&=0\;,
\end{align}
and one constraint
\begin{align}
\label{eq:constraint}
	\phi'
	+\chi'\,\frac{2\,i\,B}{r} 
	+ V_\phi\,\phi 
	+ V_\chi\,\chi
	= 0\;,
\end{align}
where primes denote derivatives with respect to~$r$.
The functions $V_\ast \equiv V_\ast(m_2^2,A,B,A',B')$ (with $\ast=\phi,\,\chi,\,\phi\phi,\,\chi\chi$) denote potentials which are independent of $\omega$. Their explicit form is given in App.~\ref{app:algebraicReduction}. In Eqs.~\eqref{eq:phi} and \eqref{eq:chi}, we have used the constraint~\eqref{eq:constraint} in order to remove mixed 1st-order terms in which both $\phi'$ and $\chi'$ appear.
\\

To explicitly solve for the constraint, we can make a general ansatz for a new master variable $\widetilde{\psi}$, i.e.,
\begin{align}
	\phi(r) &= a(r)\widetilde{\psi}(r) + b(r)\widetilde{\psi}'(r)\;,
	\\
	\chi(r) &= c(r)\widetilde{\psi}(r) + d(r)\widetilde{\psi}'(r)\;.
\end{align}
We fix the coefficients $a(r)$, $c(r)$, and $d(r)$ by demanding that the constraint equation \eqref{eq:constraint} is fulfilled. Further, we choose $b(r)=0$, for simplicity. We can then add \eqref{eq:phi} and \eqref{eq:chi} such that the resulting linear combination does not contain derivatives beyond 2nd-order. To be explicit, this requires a relative coefficient of $(-2\,i\,B/r)$. One final field redefinition, i.e.,
\begin{align}
	\widetilde{\psi}(r) = \frac{1}{r}\psi(r)\;, 
\end{align}
allows us to write the master equation in Regge-Wheeler form, i.e., results in
\begin{align}
    \frac{d^2}{dr_\ast^2}\psi(r)
	+ \left(\;
		\omega^2
		+ V(r)
	\right)\psi(r) = 0\;,
\end{align}
with a radial potential
\begin{widetext}
\begin{align}
	V(r)=\,
		-m_2^2\,A
		-\frac{(AB'+BA')}{2r}
		&-m_2^2\,\frac{
   			24\,A^2 B\,(2\,A-r\,A')\,(2\,B + r\,B')
   		}{
   		\left(
   			-4\,m_2^2\,r\,A^2\,(3\,B-1)
   			+(AB'+BA')\,(3\,r\,(AB'+BA')-4\,A)
   		\right)\,r
   		}
   		\notag\\&
   		-m_2^4\,\frac{
   			288\,A^3 B^3\,(2\,A-r\,A')^2
   		}{
   		\left(
   			-4\,m_2^2\,r\,A^2\,(3\,B-1)
   			+(AB'+BA')\,(3\,r\,(AB'+BA')-4\,A)
   		\right)^2
   		}\;.
	\label{eq:masterEq_monopole}
\end{align}    
\end{widetext}
In the Schwarzschild-limit, i.e., for $A=B=1-r_g/r$, this master equation reduces to the one previously found in \cite[Eq.(30)]{Brito:2013wya}.
\\

This master equation is the key analytical result of this work. It allows us to analyze monopole perturbations and thus long-wavelength instabilities not just on the Schwarzschild but also on the non-Schwarzschild background. We will do so in the next section.

Here, we focus on black-hole solutions in Quadratic Gravity and do not investigate horizonless objects, see e.g.~\cite{Podolsky:2019gro,Daas:2022iid}. However, the master equation \eqref{eq:masterEq_monopole} is fully general and allows to also study monopole perturbations of such non-black-hole backgrounds.

%======================================
\section{Instability of small black holes}
\label{sec:stability}
%======================================

With the different black-hole branches, cf.~Sec.~\ref{sec:bg}, and the master equation for monopole perturbations on general backgrounds, cf.~Sec.~\ref{sec:monopole-perts}, at hand, we are ready to investigate long-wavelength instabilities. We recall that a linear instability is signaled by an eigenfrequency $\omega$ with a positive imaginary part.
\\

The eigenfrequencies can be obtained by solving the master equation \eqref{eq:masterEq_monopole} with suitable boundary conditions at the horizon $r_g$ and at spatial infinity. Both boundary conditions can be obtained by an asymptotic analysis.

The horizon is a regular singular point of the master equation \eqref{eq:masterEq_monopole} and the Frobenius method can be used to extract the leading behavior. Frobenius theory tells us to expand the first-order and zeroth-order term to keep only the leading behavior, i.e., to truncate to $(r-r_g)^{-1}$ and $(r-r_g)^{-2}$, respectively, i.e.,
\ba
(r-r_g)^2 \Psi''+(r-r_g)\Psi' + \frac{\omega^2r_g^2}{b_1}\Psi=0\;.
\ea
The general solution of this frozen-coefficient equation is
\ba
\Psi (r \rightarrow r_g) \sim c_{\rm in} (r-r_g)^{\frac{-i\omega r_g}{\sqrt{b_1}}} +  c_{\rm out} (r-r_g)^{\frac{i\omega r_g}{\sqrt{b_1}}}\;.\quad
\ea
As a black hole only admits ingoing modes at the horizon, the physical solution is given by $c_{\rm out} =0$.

At spatial infinity, the master equation has an irregular singular point. Nevertheless, the leading (and subleading) behavior can be found by expanding both the first-order and zeroth-order terms, neglecting $\mathcal{O}(1/r)$ (or $\mathcal{O}(1/r^2)$), and solving the respective frozen-coefficient equation.
The general leading-order asymptotic solution is
\ba
\Psi(r_* \rightarrow +\infty) \sim c_+ e^{\sqrt{m^2-\omega^2}r_*} + c_- e^{-\sqrt{m^2-\omega^2}r_*}\; .\quad
\ea
Solutions with $c_+=0$, i.e., outgoing behavior at asymptotic infinity, describe quasinormal modes, while solutions with $c_-=0$, i.e., ingoing behavior at asymptotic infinity, describe bound states. Since we are searching for solutions with $\text{Im}(\omega r_g)<0$, the physical solution is given by the bound state, i.e., the one with $c_-=0$. 
\\

Having fixed the appropriate boundary conditions, we can compute the spectrum of bound-state perturbations. We obtain the bound-state frequencies both by a numerical forward-integration method, and by spectral methods, cf.~App.~\ref{app:spectral}.

%--------------------------------------
\subsection{Gregory-Laflamme instability of Schwarzschild spacetime}
\label{sec:GF-instab}
%--------------------------------------

%
\begin{figure}[t!]
    \centering   
    \includegraphics[height=0.63\linewidth]{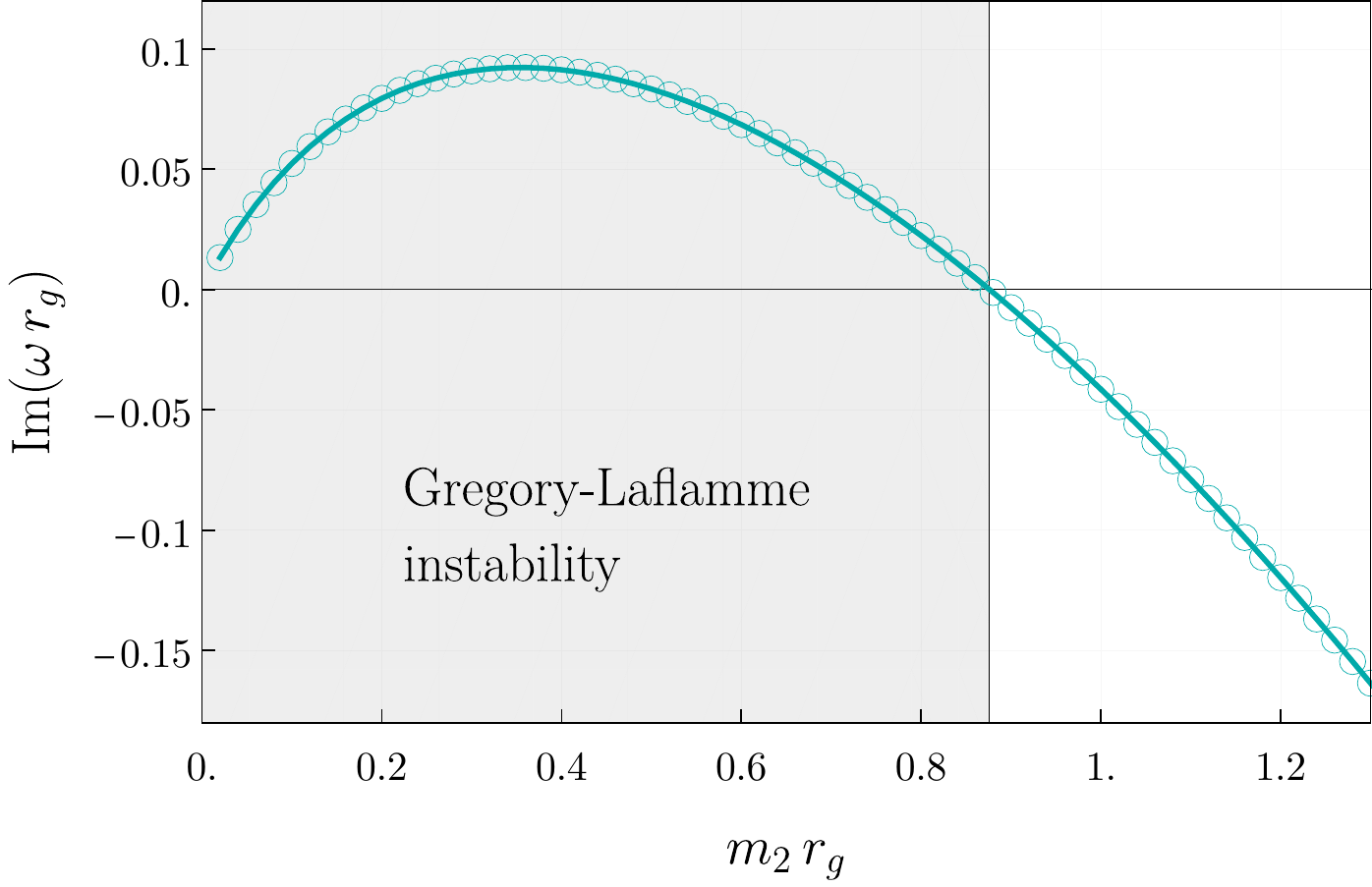}
\caption{
\label{fig:GL_instability}
	We show the imaginary part of the fundamental mode of massive spin-2 monopole perturbations on the Schwarzschild background. (The real part vanishes.) The mode is unstable ($\text{Im}(\omega r_g)>0$) for $m_2\,r_g\lesssim0.87$ and stable ($\text{Im}(\omega r_g)<0$) for $m_2\,r_g\gtrsim0.87$.
	The open cyan circles indicate results obtained by spectral methods (with negligibly small error, see App.~\ref{}).
	}
\end{figure}

The Schwarzschild black-hole background is a one-parameter family of solutions $A(r)=B(r)=1-r_g/r$ parameterized by $r_g=2M$ with $r_g$ the horizon radius and $M$ the ADM mass extracted at asymptotic infinity.
The bound-state spectrum of the massive spin-2 monopole perturbations around Schwarzschild spacetime depends on the relative size of $r_g$ and the mass $m_2$ of the massive spin-2 mode. Thus we can express results as a function of the dimensionless quantity $m_2\times r_g$.

In agreement with \cite{Brito:2013wya, Lu:2017kzi}, we find a tower of modes with $\text{Re}(\omega)=0$, cf.~Fig.~\ref{fig:newInstability}. The fundamental mode, i.e., the one with the largest imaginary part, is unstable ($\text{Im}(\omega r_g)>0$) below the branch point, i.e., for $m_2\,r_g<p\approx0.87$ and stable ($\text{Im}(\omega r_g)<0$) above the branch point, i.e., for $m_2\,r_g>p$, cf.~solid line in Fig.~\ref{fig:GL_instability}. This mode corresponds to a long-wavelength instability. In the subsequent tower of higher modes, we find no indication for further instabilities.
\\

Such a long-wavelength instability -- the Gregory-Laflamme instability -- is known from compactification of 5 (or higher) dimensional black strings to 4 dimensions~\cite{Gregory:1993vy}, cf.~\cite{Gregory:2011kh} for a pedagogical review. Indeed, the respective master equation is identical to~\eqref{eq:masterEq_monopole}, with the compactification scale $k$ taking on the role of the spin-2 mass $m_2$. 

It seems intriguing that these two a priori unrelated extensions of GR -- compactified higher dimensions on the one hand and quadratic curvature corrections on the other hand -- exhibit the exact same linear instability. Apparently, both physical scenarios reduce to the same linear degrees of freedom when the respective nonlinear dynamics is linearised around a Schwarzschild background. It is unknown whether this correspondence extends to other backgrounds or to the nonlinear dynamics.

An analytic proof for the Gregory-Laflamme instability has recently been established for $m_2\,r_g\in[3/20,8/20]$ \cite{Collingbourne:2020sfy}. Numerical results, such as the one presented here in~Fig.~\ref{fig:GL_instability}, strongly suggest that the instability is present for all 
$m_2\,r_g\in[0,0.87]$
%--------------------------------------
\subsection{Long-wavelength instability of non-Schwarzschild black holes}
\label{sec:new-instab}
%--------------------------------------

%
\begin{figure}[t!]
    \centering   
    \includegraphics[height=0.63\linewidth]{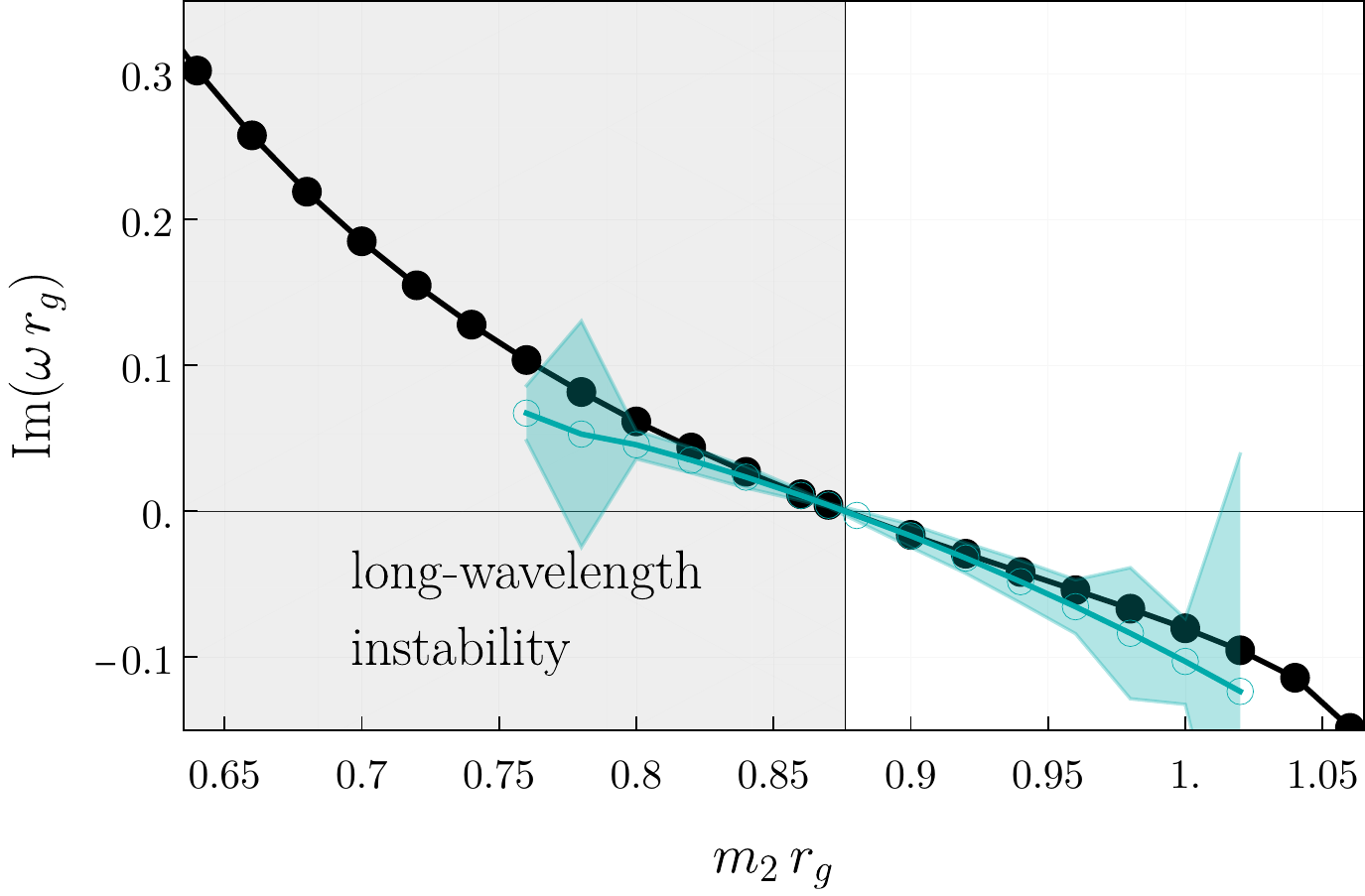}
\caption{
\label{fig:newInstability}
	We show the imaginary part of the fundamental mode of massive spin-2 monopole perturbations on the non-Schwarzschild background. (The real part vanishes.) The mode is unstable ($\text{Im}(\omega r_g)>0$) for $m_2\,r_g\lesssim0.87$ and stable ($\text{Im}(\omega r_g)<0$) for $m_2\,r_g\gtrsim0.87$. The filled black and open cyan circles indicate results obtained by forward-integration and the spectral method, respectively (see App.~\ref{app:spectralConvergence} for details on the cyan error band given for the spectral-method results).}
\end{figure}

The non-Schwarzschild black-hole branch can also be given as a one-parameter family of solutions, parameterized by the horizon radius $r_g$ but is not known in closed analytic form. Hence, we use two different methods to make sure that our conclusions about stability are converged.

The first method is fully numerical. Therein, we use the numerical black-hole solution, i.e., $A(r/r_g)$ and $B(r/r_g)$, as a background in \eqref{eq:masterEq_monopole}. The bound-state solutions are then obtained by applying the forward-integration method. We verify apparent convergence with increasing numerical precision of both the black-hole background and the forward-integration method.

The second method is to use an analytic approximation of the non-Schwarzschild branch in which the two metric functions $A(r)$ and $B(r)$ are represented by
a continued-fraction expansion parameterized by the dimensionless parameter $r_g\,m_2$. At fixed order in the continued-fraction expansion, we use a spectral method with Chebyshev polynomials to approximate the bound-state frequencies. The details of the continued-fraction expansion are reviewed in App.~\ref{app:continuedFractionApprox} and the application of spectral methods is detailed in App.~\ref{app:spectral}. As for the fully numerical solution, convergence properties are determined by a non-trivial interplay of the order $N_\text{cf}$ of the continued-fraction expansion and the order $N_\text{spec}$ of the Chebyshev polynomials.

The convergence of both methods slows down with growing distance to the branch point $p\approx 0.87$, i.e., with $\epsilon = |r_g\,m_2 - p|$. Close to the branch point, we find agreement between both methods within the respective error estimates. We are thus confident that the results for the fundamental monopole mode presented in Fig.~\ref{fig:newInstability} are converged.
\\

We find a similar picture as we found for the Schwarzschild branch. Both branches are stable for $r_g\,m_2>p$. For $r_g\,m_2<p$, both branches develop a long-wavelength instability in the fundamental monopole mode of massive spin-2 perturbations.

%======================================
\section{Discussion}
\label{sec:dicussion}
%======================================

We investigate the linear stability of spherically-symmetric black-hole solutions that arise in Quadratic Gravity, i.e., when (the action of) General Relativity is modified to include operators quadratic in the Riemann curvature.

Quadratic Gravity is known to propagate three linear degrees of freedom: the massless spin-2 graviton, a massive spin-0 mode (with mass $m_0$), and a massive spin-2 mode (with mass $m_2$), cf.~Sec.~\ref{sec:dofs}.

Among other horizonless solutions~\cite{Holdom:2002xy,Daas:2022iid,Lu:2015psa,Lu:2015cqa,Lu:2015tle,Kokkotas:2017zwt,Pravda:2016fue,Lu:2017kzi,Podolsky:2019gro}, Quadratic Gravity exhibits two branches of static, spherically-symmetric, and asymptotically flat black hole solutions, cf.~Sec.~\ref{sec:bg} and Fig.~\ref{fig:syopsis}. Each branch represents a one-parameter family of black holes, parameterized by a dimensionless parameter $r_g\, m_2$. The two branches intersect at a branch point $r_g\, m_2\equiv p \approx 0.87$.

%--------------------------------------
\subsection{Key result: Long-wavelength instability\\for small black holes in both branches}
%--------------------------------------

We have uncovered a long-wavelength instability in the non-Schwarzschild branch of spherically symmetric black holes in Quadratic Gravity. This instability complements the Gregory-Laflamme instability~\cite{Gregory:1993vy,Gregory:2011kh,Collingbourne:2020sfy} of the Schwarzschild branch to form a lower bound for the horizon radius $r_g>p/m_2$ of stable (spherically-symmetric) black holes in Quadratic Gravity.
\\

To obtain this result, we work in the Einstein frame and derive the covariant equations of motion for general linear perturbations. We explicitly show that the massive spin-0 mode decouples on any `Ricci-scalar-flat' ($R=0$) background and can thus be treated separately, cf.~Sec.~\ref{sec:perts}. The massless and the massive spin-2 mode remain coupled unless the background is a vacuum solution to General Relativity, i.e., is Ricci-flat ($R_{\mu\nu}=0$). Nevertheless, we can derive a Regge-Wheeler type master equation \eqref{eq:masterEq_monopole} for the monopole perturbation about an arbitrary static spherically-symmetric background. The lowest lying modes in the respective bound-state spectrum reveal both instabilities. 

On the Schwarzschild background, the master equation reduces to a well-known result~\cite{Brito:2013wya, Lu:2017kzi} and we recover the Gregory-Laflamme instability~\cite{Gregory:1993vy,Gregory:2011kh,Collingbourne:2020sfy} for small black holes with horizon radius $r_g\,m_2<p$, cf.~Sec.~\ref{sec:GF-instab}. 

On the non-Schwarzschild background, we uncover a similar long-wavelength instability for small black holes, the onset of which is, once again, set by the branch point, i.e., the instability occurs for $r_g\,m_2<p$, cf.~Sec.~\ref{sec:new-instab}.

%--------------------------------------
\subsection{Effective field theory of General Relativity}
%--------------------------------------

Quadratic-curvature operators are also present in the effective field theory (EFT) of General Relativity, the action of which, up to leading EFT corrections (and for negligible cosmological constant), takes the same form as Eq.~\eqref{eq:QG-action}. Within this EFT, one has to ensure that both the background solution and the linear stability analysis do not extrapolate the EFT beyond its regime of validity.
\\

In a weakly coupled EFT, the masses of the massive spin-0 and spin-2 degrees of freedom are comparable to or lie beyond the EFT cut-off scale: Specifically, the cut-off scale is given by $m_0 \sim m_2$ \cite{deRham:2019ctd}, and hence the massive spin-0 and spin-2 fields cannot be excited within the validity of the EFT.

In a generic (potentially strongly coupled) UV completion, we expect $m_0 \sim m_2 \sim \mpl$, in which case the mass scales associated with the massive spin-0 and spin-2 degrees of freedom are either above the cut-off scale or the EFT becomes strongly coupled as the energy scale approaches $\mpl$.
In the latter case, the linear analysis breaks down and one can potentially no longer neglect terms of yet higher order (cubic, quartic, and so on) in curvature.
\\

The EFT cutoff scale also has implications for the validity of the background solutions.
In the Schwarzschild branch, the largest curvature scales (in the black-hole exterior) occur at the horizon and are uniquely determined by the Kretschmann scalar $R_{\mu\nu\rho\sigma}R^{\mu\nu\rho\sigma}/m_2^4 = 12/(r_g^4\, m_2^4)$: The smaller the black hole, the larger its horizon curvature and, in particular, $R_{\mu\nu\rho\sigma}R^{\mu\nu\rho\sigma}/m_2^4 = 12/p^4$ at the branch point. Large Schwarzschild black holes above the branch point are thus within the validity of the EFT but for small black holes close to and below the branch point the EFT presumably breaks down.

\begin{figure}[t!]
    \centering   
    \includegraphics[height=0.63\linewidth]{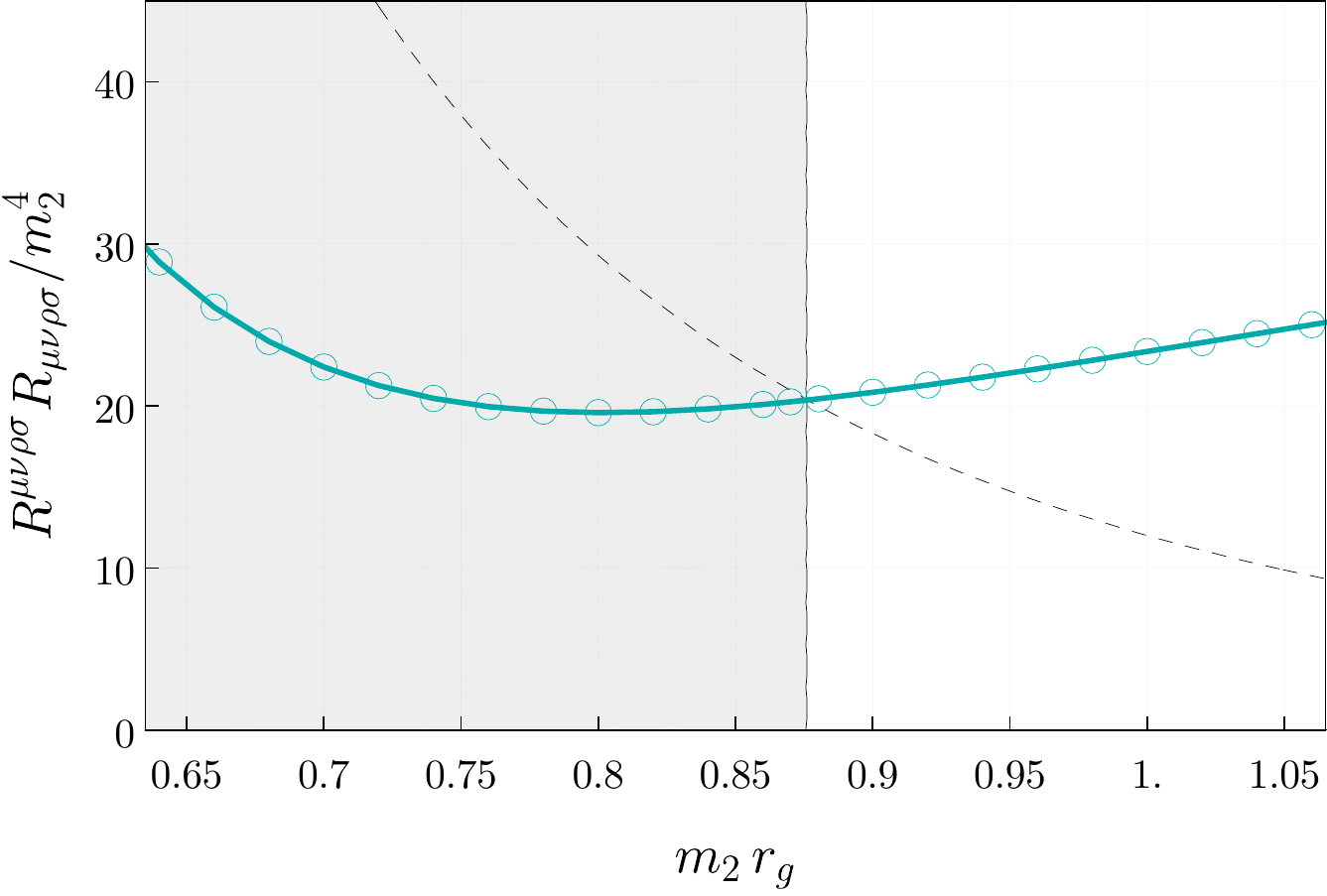}
\caption{\label{fig:R2h}
We show the Kretschmann scalar $R_{\mu\nu\rho\sigma}R^{\mu\nu\rho\sigma}/m_2^4$ evaluated at the black hole horizon. The Kretschmann scalar of the Schwarzschild branch (thin dashed) decreases monotonically as the horizon increases, hence large Schwarzschild black holes are within the validity of the EFT. In contrast, the Kretschmann scalar of the non-Schwarzschild branch (thick cyan) starts to increase as the horizon grows for $m_2\,r_g>p$. Hence, we expect that large non-Schwarzschild black holes are not within the validity of the EFT.}
\end{figure}

In contrast,
we expect that black holes in the non-Schwarzschild branch are always outside the validity of the EFT: On the one hand, the EFT is only valid as long as quadratic-curvature terms contribute perturbatively. On the other hand, we find that the Kretschmann scalar, evaluated at the horizon of the non-Schwarzschild branch, increases with growing horizon size. We thus expect that even large non-Schwarzschild black holes exhibit curvature scales beyond the EFT cut-off scale, cf.~Fig.~\ref{fig:R2h}.

%--------------------------------------
\subsection{A scenario for remnants}
%--------------------------------------

Our results raise an interesting question concerning the overall fate of black holes in Quadratic Gravity. Both black-hole branches develop a linear long-wavelength instability once the horizon radius drops below $r_g<p/m_2$ which implies a lower bound on the horizon radius of stable black holes. These uncovered long-wavelength instabilities are driven purely by classical perturbations. At the same time, semiclassical matter fluctuations (obtained in quantum field theory on curved spacetime) lead to Hawking radiation and thus to a decreasing horizon radius -- at least in the Schwarzschild branch. The competition of both instabilities could thus lead to a mechanism that stabilizes black holes at the branch point, cf.~Fig.~\ref{fig:syopsis}, and could thus lead to stable remnants with a characteristic horizon radius of $r_g = p/m_2$.
\\

Semiclassical black-hole perturbations of quantized matter fields in curved spacetime lead to Hawking evaporation~\cite{Hawking:1974rv}.
A large (i.e., $m_2\,r_g\gg p$) Schwarzschild black hole will evaporate and is thus driven towards the branch point.
The timescale of Hawking evaporation $t_\text{ev}$ is set by $t_\text{ev}/\tpl \sim M^3/\mpl^3$ with $M$ the black hole mass and $\mpl$ ($t_\text{Pl}$) the Planck mass (Planck time). The underlying semiclassical approximation neglects backreaction and thus breaks down as $M\rightarrow\mpl$. For all observed astrophysical black holes, $M>M_\odot$ and thus the evaporation timescale is much longer than the observed age of the universe. However, there exists an intermediate regime of black-hole masses $M$, in which $M_\odot\gg M\gg\mpl$: In this regime the semiclassical approximation is valid and the timescale is observable. For instance, black holes of mass $M\sim10^{11}\,\text{kg}$ would have an evaporation timescale of roughly one year. 

In summary, Hawking radiation drives Schwarzschild black holes from larger to smaller horizon radii.
In contrast, we have seen that quadratic-curvature corrections to General Relativity lead to classical long-wavelength instabilities that destabilize small black holes below a critical horizon radius set by the mass scale $m_2$ associated with the Quadratic Gravity spin-2 degree of freedom.

Whether or not the classical instability can counteract semiclassical Hawking evaporation depends on the respective timescales of both instabilities. On the one hand, the time scale for developing the Gregory-Laflamme instability can be estimated by $t_{\rm GL}\sim 1/{\rm Im}[\omega]$. On the other hand, the time scale for complete evaporation is $t_{\rm ev} \approx 5120\,G^2 M^3$~\cite{Hawking:1975vcx}. Thus, we find that $t_{\rm GL}$ is larger than $t_{\rm ev}$ if $m_2 \gtrsim 36 \mpl$. In turn, we expect that the Gregory-Laflamme instability occurs before complete evaporation (down to a Planck-sized mass) of the black hole if $m_2 \lesssim 36 \mpl$. This also implies that black holes at the branch point have a mass of at least $M/\mpl=4\pi\,p\,\mpl/m_2 \gtrsim 144\pi\,p \gg 1$ such that we can in fact trust the semiclassical calculation throughout the whole process.
\\

Overall, this implies that Schwarzschild black holes in Quadratic Gravity evaporate until they reach a horizon radius of $r_g = p/m_2$, at which point the Gregory-Laflamme instability occurs. The crucial remaining question is the fate of this classical instability. In spherical symmetry, we see, a priori, four remaining options, cf.~Fig.~\ref{fig:syopsis}: (i) the black hole continues its decay to smaller horizon radii in the non-Schwarzschild branch; (ii) the black hole decays to larger horizon radii in the non-Schwarzschild branch; (iii) the theory develops a runaway, or finally (iv) the competing instabilities balance out to form stable remnants with characteristic horizon radius $r_g = p/m_2$.

Due to the uncovered long-wavelength instability in the non-Schwarzschild branch, we can now exclude option (i).
Thermodynamic arguments~\cite{Fan:2014ala, Goldstein:2017rxn, Lu:2017kzi}, following the Wald formalism \cite{Wald:1993nt, Iyer:1994ys}, suggest that Hawking radiation will also shrink the horizon size of black holes in the non-Schwarzschild branch. While an explicit semiclassical calculation remains outstanding, this suggests that also option (ii) is excluded.
Finally, an understanding of option (iii) presumably requires the full non-linear dynamics of Quadratic Gravity.

%--------------------------------------
\subsection{Outlook}
%--------------------------------------

Regarding the ultimate fate of spherically-symmetric black holes in Quadratic Gravity, we can thus identify two key open questions: 
\begin{itemize}
    \item
    Do semiclassical fluctuations drive large black holes towards the branch point, as suggested by thermodynamic considerations~\cite{Fan:2014ala, Goldstein:2017rxn, Lu:2017kzi}?
    \item 
    Does the full nonlinear dynamics of Quadratic Gravity (see~\cite{Noakes:1983, Held:2021pht} for an approach to well-posed numerical evolution) avoid runaways?
\end{itemize}
If both of these questions can be answered in the affirmative, we may expect the formation of stable remnants -- at least in spherical symmetry.
\\

Little is known beyond spherical symmetry and alternative branches of axisymmetric solutions have not yet been found\footnote{Alternative axisymmetric black-hole branches need not be circular~\cite{Xie:2021bur,Delaporte:2022acp}. In fact, an application of the Janis-Newman complexification~\cite{Newman:1965tw} -- generalizing static and spherically symmetric solutions to axisymmetric and stationary candidate solutions -- is known to only generate circular (and moreover algebraically special) spacetimes. While successfully generalizing Schwarzschild (Reissner-Nordstr\"om) spacetime to Kerr (Kerr-Newman) spacetime, all of which are circular and algebraically special~\cite{Drake:1998gf}, the Janis-Newman complexification will presumably fail in Quadratic Gravity.}.
Since vacuum solutions to GR are Ricci flat, Kerr spacetime -- just like Schwarzschild spacetime -- is also a solution to Quadratic Gravity. For slowly spinning Kerr spacetime the superradiance phenomenon has been effectively explored in Quadratic Gravity~\cite{Brito:2013wya,Brito:2020lup}. A full analysis (akin to the Teukolsky equation~\cite{Teukolsky:1973ha} in GR) remains outstanding. We plan to investigate the case of axisymmetry in future work.

%======================================
\begin{acknowledgments}
We thank Claudia de Rham for many helpful conversations and comments. The work of AH at Imperial College London was supported by the Royal Society International Newton Fellowship NIF{\textbackslash}R1{\textbackslash}191008. The work leading to this publication was supported by the PRIME programme of the German Academic Exchange Service (DAAD) with funds from the German Federal Ministry of Education and Research (BMBF). The work of JZ at Imperial College London was supported by the European Union's Horizon 2020 Research Council grant 724659 MassiveCosmo ERC-2016-COG. J.Z. is also supported by the scientific research starting grants from University of Chinese Academy of Sciences (grant No.~118900M061) and the Fundamental Research Funds for the Central Universities (grant No.~E2EG6602X2 and grant No.~E2ET0209X2).

\end{acknowledgments}

%======================================
\appendix
%======================================

%======================================
\section{The horizonless limit of the non-Schwarzschild solution} 
\label{app:singularity}
%======================================
In this appendix, we investigate the horizonless limit of the non-Schwarzschild solution. It is convenient to work with an alternative metric form  \cite{Pravda:2016fue}, 
\ba
{\rm d} s^2 = \Omega^2(\bar{r})\left[\,{\rm d} \theta^2+\sin^2\theta\, {\rm d}\phi^2 -2\,{\rm d} u\, {\rm d} r+{\cal H}(r)\, {\rm d} u^2 \,\right], \nonumber \\
\ea 
which relates to \eqref{eq:metric} via the transformation
\ba
r = \Omega(\bar{r}), \quad t=u - \int {\cal H}^{-1}  {\rm d} r,
\ea
with
\ba
 A(r) = -\Omega^2\, {\cal H} \,,\qquad  B(r) = -(\Omega'/\Omega)^2\, {\cal H} \,.
\ea
The horizon is located at ${\cal H}(\bar{r}_g)=0$. To see the curvature singularity in the horizonless limit of the non-Schwarzschild solution, we consider the Bach curvature tensor
\ba
B_{\mu\nu} \equiv \left(\nabla^{\rho}\nabla{\sigma} + \frac12 R^{\rho\sigma}\right)C_{\mu\rho\nu\sigma}\, ,
\ea
and evaluate the Bach scalar curvature invariant at the horizon (for example see Ref.~\cite{Svarc:2018coe}),
\ba
B_{\mu\nu}B^{\mu\nu} &=& \frac{1}{4} \left(\frac{b}{4 m_2^2}\right)^2 \Omega^{-4} (\bar{r}_g) \nonumber \\ 
&=& \frac{1}{4} \left(\frac{b}{4 m_2^2}\right)^2 r_g^{-4} \label{eq:B2}
\ea
where $b$ is the non-zero parameter that parameterizes the non-Schwarzschild solutions. In the horizonless limit, i.e., as $r_g \rightarrow 0$, the Bach scalar curvature diverges, i.e., $B_{\mu\nu}B^{\mu\nu}\rightarrow\infty$, and thus the non-Schwarzschild solution turns into a naked curvature singularity.

%======================================
\section{Algebraic reduction of the monopole perturbations}
\label{app:algebraicReduction}
%======================================

Prior to gauge fixing and any reduction, the monopole ($\ell=0$) perturbations in spherically-symmetric Quadratic Gravity are described by 8 polar mode functions, i.e., by $H_{0,1,2}$, $\mathcal{K}$, $F_{0,1,2}$, and $\mathcal{M}$, cf.~\eqref{eq:evenfab} and~\eqref{eq:evenhab}. The respective metric (Eq.~\eqref{eq:linear-eom-metric}) and massive spin-2 (Eq.~\eqref{eq:linear-eom-ghost}) equations of motion, i.e., $\mathcal{F}=0$ and $\mathcal{H}=0$ reduce to 4 non-trivial equations each, i.e., to
\begin{align}
	\mathcal{F}_{tt}&=0\,,\quad
	\mathcal{F}_{tr}=0\,,\quad
	\mathcal{F}_{rr}=0\,,\quad
	\mathcal{F}_{\theta\theta}=0\,,\quad
	\\
	\mathcal{H}_{tt}&=0\,,\quad
	\mathcal{H}_{tr}=0\,,\quad
	\mathcal{H}_{rr}=0\,,\quad
	\mathcal{H}_{\theta\theta}=0\;.
\end{align}
In spherical symmetry, the other components are either trivial or equivalent to one of the above equations.

The constraints, i.e., $D_a f^{ab}=0$ and $f=0$, correspond to three non-trivial equations, i.e.
\begin{align}
	\mathcal{C}^t\equiv D_a f^{at}=0\,,\quad
	\mathcal{C}^r\equiv D_a f^{ar}=0\,,\quad
	f=0\,.
\end{align}
As written in the main text, we choose a gauge in which
\begin{align}
	\mathcal{K}=H_{1}=0\;.
\end{align}
Further, we algebraically solve $\mathcal{C}^t=0$ and $f=0$ to for $F_0$ and $F_2$. Finally, we algebraically solve two of the metric-perturbation equations, i.e., $\mathcal{F}_{tr}=0$ and $\mathcal{F}_{rr}=0$, for $H_2$ and $H_0'(r)$. (We can solve for $H_0'(r)$ directly because none of the equations contains $H_0(r)$.)
\\

Overall, this leaves us with two massive spin-2 perturbations, i.e., $F_1$ and $\mathcal{M}$, and six non-trivial equations, i.e., 
$\mathcal{H}_{\theta\theta}=0$,
$\mathcal{F}_{tt}=0$,
$\mathcal{F}_{tr}=0$,
$\mathcal{F}_{rr}=0$,
$\mathcal{F}_{\theta\theta}=0$, and
$\mathcal{C}^r=0$.
Defining,
\begin{align}
	\phi(r) &= -2\omega\mathcal{M}(r)\;,
	\\
	\chi(r) &= F_1(r)\;,
\end{align}
these six equations can be algebraically reduced to one 2nd-order equation for $\phi$ (0th-order in $\chi$), one 2nd-order equation for $\chi$ (0th-order in $\phi$), and one 1st-order constraint. The explicit linear combinations which achieve this reduction read
\begin{widetext}
\begin{align}
	0&=-\frac{2\,\omega}{B\,r^2}\,\mathcal{F}_{\theta\theta} + \left(
		\frac{A'}{A}-\frac{B'}{A}+\frac{2}{r}
	\right)\left(
		\omega\,\mathcal{C}^r
		-i\,\mathcal{F}_{tr}
	\right)
	\notag\\
	&\equiv
	-i\,\frac{d}{dr}(\mathcal{F}_{tr})
	+\omega\,\mathcal{F}_{rr}
	+\left(
		\frac{A'}{A}
		-\frac{1}{r}\left(1+\frac{1}{B}\right)
	\right)\omega\,\mathcal{C}^r
	-\frac{1}{2} \left(
		-\frac{3 A'}{A}
		+\frac{B'}{A}
		-\frac{8}{r}
	\right)\left(
		\omega\,\mathcal{C}^r
		-i\,\mathcal{F}_{tr}
	\right)
	\;,
	\\
	0&=\frac{1}{B}\,\mathcal{F}_{tr}
	\;,
	\\
	0&=
	\omega\,\mathcal{C}^r-i\,\mathcal{F}_{tr}
	\equiv
	\frac{2\,\omega}{M^2r^3}\mathcal{H}_{\theta\theta}
	-i\,\mathcal{F}_{tr}
	\equiv
	\frac{\omega\,r}{A\left(r\,B'+2\,B-2\right)}\left[
		\mathcal{F}_{tt}
		-A\,B\,\mathcal{F}_{rr}
		-\frac{2\,A}{r^2}\,\mathcal{F}_{\theta\theta}
		-B\,A'\,\mathcal{C}^r
	\right]-i\,\mathcal{F}_{tr}
	\;.
	\end{align}
The resulting equations are given in the main text, cf.~Eqs.~\eqref{eq:phi}-\eqref{eq:constraint}. The explicit expressions for the potentials read
\begingroup
\allowdisplaybreaks
\begin{align}
	\mathcal{V}_{\phi\phi} =&
	-\frac{A (3 B-1) m_2^2}{B^2 \left(2 A-r A'\right)}
	-\frac{1}{4 A^2 B^2 r^2 \left(r A'-2 A\right)}\Bigg[
		A^3 \left(r B'+4 B\right) \left(3 r B'+4 B-4\right)
		\notag\\&\quad\quad
		+2 B^2 r^3 \left(A'\right)^3-A B r^2 \left(A'\right)^2 \left(2 r B'+17 B\right)
		+2 A^2 B r A' \left(r B'+12B+2\right)
   \Bigg]
   \;,
	\\[1em]
	\mathcal{V}_{\phi\chi} =&
	\frac{2 i m_2^2}{B^2}\left(
		\frac{A (3 B-1) B'}{r A'-2 A}
		+\frac{B}{r}
	\right)
	\notag\\&
	-\frac{i}{2 A^2 B^2 r^3 \left(r A'-2 A\right)}\Bigg[
		B^3 r^3\left(A'\right)^3
		+A B^2 r^2 \left(A'\right)^2 \left(-7 r B'-6 B+4\right)
		+A^2 B r^2 A' B' \left(3 r B'+8 B\right)
		\notag\\&\quad\quad
		A^3 \left(
			16 (B-1) B^2
			+3 r^3 \left(B'\right)^3
			+2 (3 B-2) r^2 \left(B'\right)^2
			+8 B (3 B-1) r B'
		\right)
   \Bigg]
	\;,
	\\[1em]
	\mathcal{V}_{\chi\phi} =&
	\frac{i \left(A B'-3 B A'\right)}{2 A B^2}
	\;,
	\\[1em]
	\mathcal{V}_{\chi\chi} =&
	-\frac{m_2^2}{A^2} \left(
		\frac{3 A (3 A-1)}{2 A-r A'}
		-2 A
	\right)
	+\frac{1}{4 A^2 B^2 r^2 \left(r A'-2 A\right)}\Bigg[
		A^3 \left(
			r B' \left(12-5 r B'\right)
			+4 B \left(4-3 r B'\right)
		\right)
		\notag\\&\quad\quad
		+4 B^2 r^3 \left(A'\right)^3
		+A B r^2 \left(A'\right)^2 \left(2 r B'+5
   B\right)
   		-2 A^2 r A' \left(
   			22 B^2+r^2 \left(B'\right)^2
   			+B \left(8 r B'-2\right)
   		\right)
	\Bigg]
	\;,
	\\[1em]
	\mathcal{V}_{\phi} =&
	\frac{3}{r}-\frac{A'}{2 A}
	\;,
	\\[1em]
	\mathcal{V}_{\chi} =&
	-\frac{i A (3 B-1) m_2^2}{B \left(r A'-2 A\right)}
	-\frac{i}{4 A B r^2
   \left(r A'-2 A\right)}
	\Bigg[
		A^2 \left(32 B^2-3 r^2 \left(B'\right)^2+4 (4 B+1) r B'\right)
		\notag\\&\quad\quad
		+B^2 r^2 \left(A'\right)^2-2 A B r A' \left(7 r B'+12 B-2\right)
	\Bigg]\;.
\end{align}
\endgroup
\end{widetext}

%======================================
\section{Continued-fraction expansion of the non-Schwarzschild background}
\label{app:continuedFractionApprox}
%======================================

In this appendix, we collect the relevant details of the continued-fraction expansion \cite{Kokkotas:2017zwt} (based on the general framework in \cite{Rezzolla:2014mua}) of the non-GR background solution \cite{Lu:2015cqa, Lu:2015psa} of Quadratic Gravity.
\\

Defining a dimensionless compact coordinate $x = 1-r_0/r$, where $r_0$ is the horizon of the non-GR black hole, the two metric functions $A(r)$ and $B(r)$ in Eq.~\eqref{eq:general-sph-sym-BG} are expanded as
\begin{align}
	A(r) &= x\,\widetilde{A}(x)\;,
	\quad\quad\quad
	A(r)/B(r) = \widetilde{B}(x)^2\;,	
\end{align}
with
\begin{align}
    \widetilde{A}(x)&=1-\epsilon (1-x)+(a_0-\epsilon)(1-x)^2+\widehat{A}(x)(1-x)^3\;,
	\notag\\
	\widetilde{B}(x)&=1+b_0(1-x)+\widehat{B}(x)(1-x)^2\;,
\end{align}
and
\begin{align}
	\widehat{A}(x)&=
	\frac{a_1}{
		\displaystyle 1+\frac{\displaystyle a_2\,x}{
			\displaystyle 1+\frac{\displaystyle a_3,x}{
				\displaystyle 1+\frac{\displaystyle a_4\,x}{
					\displaystyle 1+\ldots
				}
			}
		}
	}\;,
    \notag\\
	\widehat{B}(x)&=
	\frac{b_1}{
		\displaystyle 1+\frac{\displaystyle b_2\,x}{
			\displaystyle 1+\frac{\displaystyle b_3\,x}{
				\displaystyle 1+\frac{\displaystyle b_4\,x}{
					\displaystyle 1+\ldots
				}
			}
		}
	}\;.
\end{align}
The background equations of motion imply $a_0 = b_0 = 0$.
\\

At 2nd order, the other expansion coefficients are given by
\begin{align}
	\epsilon&\approx\frac{1054 - 1203\,p}{326}\;,
	\notag\\
	a_1&\approx\frac{1054 - 1203\,p}{556}\;,
	\quad\quad\quad\quad\;\;\,
	b_1\approx-\frac{1054 - 1203\,p}{1881}\;,
	\notag\\
	a_2&\approx-\frac{18 - 17\,p}{11}\;,
	\quad\quad\quad\quad\quad\quad
	b_2\approx-\frac{2+p}{4}\;,
\end{align}
setting $a_{i>2}=0$ and $b_{i>2}=0$.
At 4th order, the other expansion coefficients read
\begin{align}
	\epsilon&\approx(1054 - 1203 p)\left(\frac{3}{1271} + \frac{p}{1529}\right)\;,
	\notag\\
	a_1&\approx(1054 - 1203 p)\left(\frac{7}{1746}-\frac{5 p}{2421}\right)\;,
	\notag\\
	b_1&\approx(1054 - 1203 p)\left(\frac{p}{1465}-\frac{2}{1585}\right)\;,
\end{align}
\begin{align}
	a_2&\approx\frac{6 p^2}{17}+\frac{5 p}{6}-\frac{131}{102}\;,
	\notag\\
	b_2&\approx\frac{81 p^2}{242}-\frac{109 p}{118}-\frac{16}{89}\;,
\end{align}
\begin{align}
	a_3&\approx\frac{\dfrac{9921 p^2}{31}-385 p+\dfrac{4857}{29}}{237-223 p}\;,
	\notag\\
	b_3&\approx-\frac{2 p}{57}+\frac{29}{56}\;,
\end{align}
\begin{align}
	a_4&\approx\frac{\dfrac{9 p^2}{14}+\dfrac{3149 p}{42}-\dfrac{2803}{14}}{237-223 p}\;,
	\notag\\
	b_4&\approx\frac{13 p}{95}-\frac{121}{98}\;,
\end{align}
again setting $a_{i>4}=0$ and $b_{i>4}=0$.

%======================================
\section{Spectral Methods}
\label{app:spectral}
%======================================

 %
\begin{figure}
    \centering   
    \includegraphics[width=\linewidth]{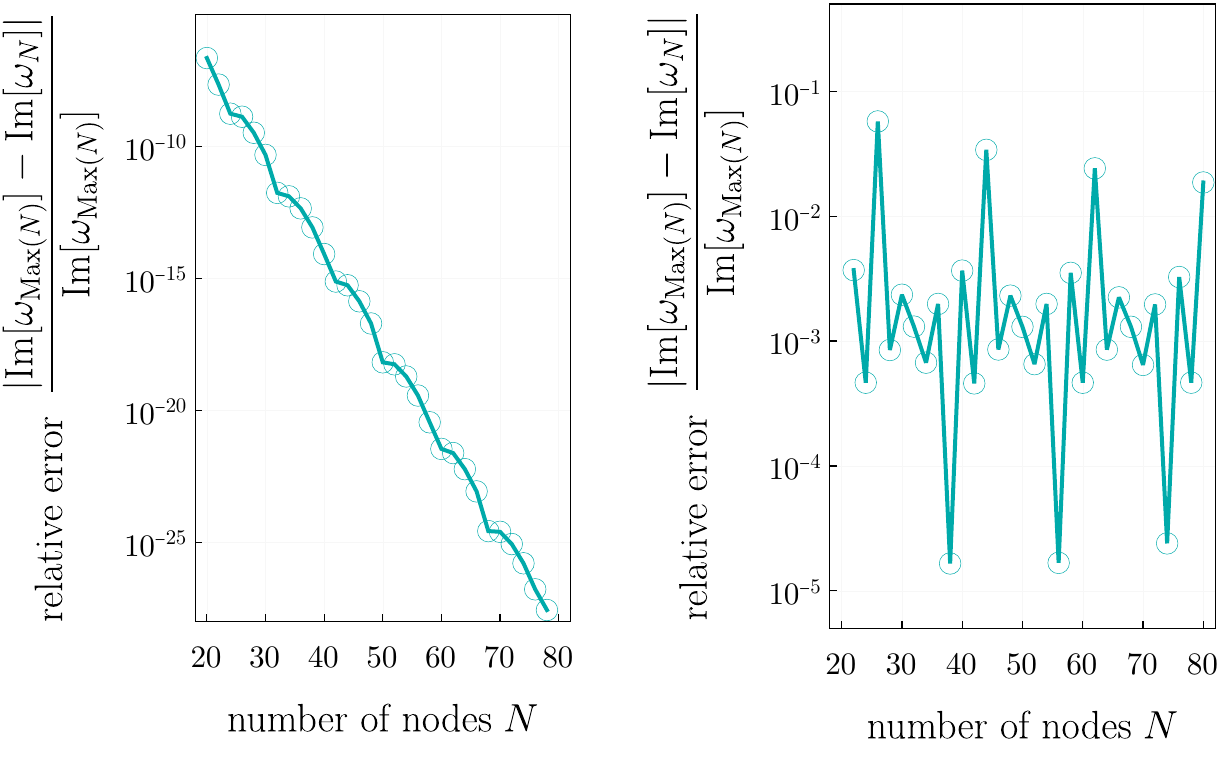}
\caption{
\label{fig:convergence}
	Exemplary convergence of the fundamental massive spin-2 monopole mode with growing number $N$ of Chebyshev nodes. More specifically, we plot the absolute difference between the current-order $N$ and the highest-order $Max(N)=80$ result normalized by the $Max(N)=80$ result. In the left-hand panel, we show the (fast) convergence for the exact Schwarzschild background (at a randomly selected point within the domain in Fig.~\ref{fig:GL_instability}). In the right-hand panel, we show the (slow) convergence for the 4th-order continued-fraction approximation of the non-Schwarzschild background (at $m_2\,r_g = 0.9$).}
\end{figure}
\begin{figure}
    \centering   
    \includegraphics[width=\linewidth]{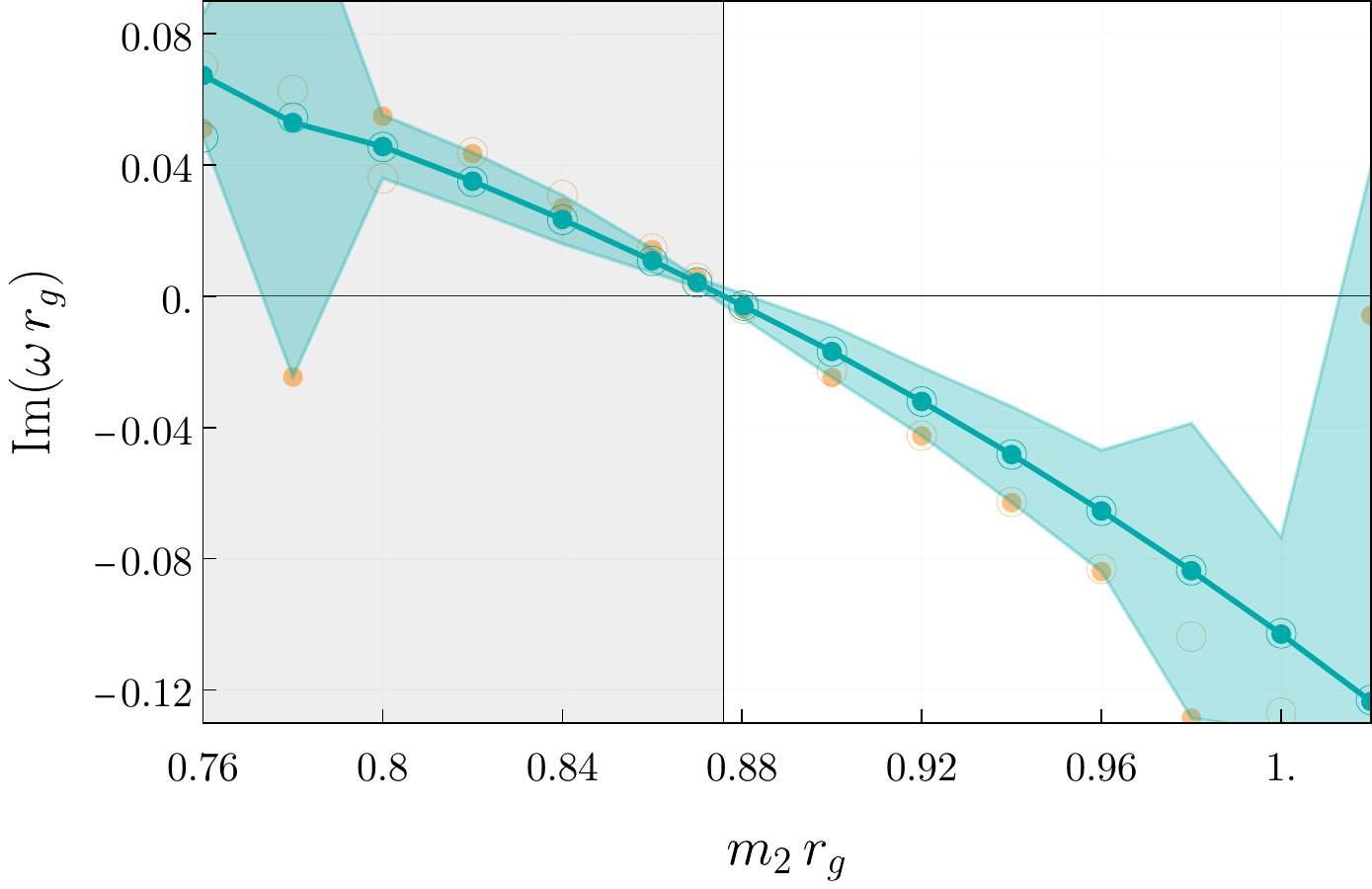}
\caption{
\label{fig:spectral_nonSchw}
	Imaginary part of the fundamental mode of massive spin-2 monopole perturbations obtained by spectral methods (see App.~\ref{app:spectral}) on the continued-fraction approximation (see App.~\ref{app:continuedFractionApprox}) of the non-Schwarzschild background. (The real part vanishes.) The open circles and dots indicate results obtained with $N=80$ and $N=20$ Chebyshev nodes, respectively. In both cases, the cyan and light orange points indicate results obtained on a 4th-order and 2nd-order continued-fraction approximation of the background, respectively.}
\end{figure}
In addition to a generic shooting method on a fully numerical background, we also use spectral methods with Chebyshev collation points and an analytical continued-fraction expansion of the background (cf.~App.~\ref{app:continuedFractionApprox}) to determine the relevant eigenfrequencies of the master equation~\eqref{eq:masterEq_monopole} derived in the main text.

For any real-valued function $f(x)$ defined on a finite interval $x\in [a,b]$ and any $\epsilon\in\mathbb{R}$, there exists a polynomial $p$ of degree $N$ for which $|f(x)-p(x)|<\epsilon$ (Weierstrass approximation theorem).
Hence, there must exist a series of polynomials that converges to the eigenfunctions (and the respective frequencies which converge to the eigenfrequencies) of an ordinary differential equation such as Eq.~\ref{eq:masterEq_monopole}.
The crux is to find an appropriate series of collation points $x_1,\,\dots,\,x_N$, at which $p(x_N)=f(x_N)$ to ensure uniform convergence: For generic choices of collation points, polynomials of growing $N$ rapidly diverge at the edges of the interval (Runge's phenomenon). Chebyshev polynomials are special in that this choice of collation points minimizes Runge's phenomenon.

Before applying spectral methods, we factor out the desired boundary behavior, both at the horizon and at asymptotic infinity. The asymptotic analysis, cf.~Sec.~\ref{sec:stability}, results in
\begin{align}
\label{eq:boundary-conditions-GRBG}
	\psi_\text{Schw}(r) =& 
	(r_g-r)^{-i\,r_g\,\omega\,\rho}\times
	r^{i\,r_g\,\omega\,\rho + \frac{r_g(m_2^2 - 2\omega^2)}{2 q}\,\delta}
	\notag\\
	&\times
	e^{q\,r}\times
	R(r)\,r\;,
\end{align}
with the sign of $q=\pm\sqrt{m_2^2 - \omega^2}$ differentiating between ingoing (bound-state) and outgoing (quasi-normal) modes at asymptotic infinity. Moreover, $\rho=\rho(r_gm_2)$ and $\delta=\delta(r_gm_2)$ are constants which depend on the specified background. 

For Schwarzschild spacetime, $\rho\equiv\delta\equiv1$. For the non-Schwarzschild background, $\rho$ and $\delta$ depend on the order of the continued-fraction expansion (cf. App.~\ref{app:continuedFractionApprox}). At second order in $p=r_gm_2$, we find
\begin{align}
	\rho(p) &= \frac{90628 (1203 p+827)}{1881 (472779 p-323594)}\;,
	\\
	\delta(p) &= \frac{3}{326} (460 - 401p)\;.
\end{align}
At fourth order in $p=r_gm_2$, we find
\begin{widetext}
\begin{align}
	\rho(p) &= -\frac{912745194966 \left(381351 p^2-1039076 p+153239\right)}{464405 \left(3704000354214 p^2-2463952425415 p+228230924900\right)}\;,
	\\
	\delta(p) &= \frac{143}{41} -\frac{4178527 p}{1943359} -\frac{1203 p^2}{1529}\;.
\end{align}
\end{widetext}
We recast the radial variable,
\begin{align}
	\xi = \frac{r- 2\sqrt{r\,r_g}}{r},
\end{align}
such that $\xi\stackrel{r\rightarrow r_g}{\longrightarrow}-1$ and $\xi\stackrel{r\rightarrow\infty}{\longrightarrow}1$. The master equation can then be written in standard form
\begin{align}
\label{eq:master-xi}
	\left(
	    \frac{\text{d}^2}{\text{d}\xi^2} 
	    + C_1(\omega,\, \xi) \frac{\text{d}}{\text{d}\xi} 
	    + C_2(\omega,\, \xi)
	\right) R(\xi) = 0 \,,
\end{align}
where $C_i$ are functions of the radial variable $\xi$ and the frequency $\omega$.
The Chebyshev polynomials (of order $N$) approximating $R(\xi)$ are defined as
\begin{align}
    R_N(\xi) \equiv \sum_{k=0}^{N} R(\xi_k) p_k(\xi) \,,
\end{align}
with $p_k(\xi_n) \equiv \delta_{nk}$ at the Chebyshev nodes
\begin{align}
	\xi_n \equiv \cos \left(\frac{\pi(2n+1)}{2N+2}\right), 
	\quad
	{\rm with}
	\quad
	n = 0,1, \ldots, N\,.
\end{align}
To be explicit, the polynomials $p_k(\xi)$ are defined as
\begin{align}
    p_k(\xi) \equiv \frac{p(\xi)}{(\xi - \xi_k)}\,q_k\;,
\end{align}
in terms of the $(N+1)$-order node polynomial
\begin{align}
    p(\xi) \equiv \prod_{k=0}^N (\xi - \xi_k)
\end{align}
and the weights
\begin{align}
    q_k \equiv \left(\left.\frac{\text{d}p(\xi)}{\text{d}\xi}\right|_{\xi=\xi_k}\right)^{-1}\;.
\end{align}
At each order $N$, the ODE is then approximated by an algebraic system of $N$ equations, i.e.,
\begin{align}
\label{eq:master-matrix}
	\sum_{k=0}^N {\cal M}_{nk}(\omega) R(\xi_k) =0 \,,
\end{align}
with
\begin{align}
	{\cal M}_{nk}(\omega) \equiv p_k''(\xi_n) + C_1(\omega, \xi_n) p_k'(\xi_n) +C_2(\omega,\xi_n)\delta_{nk} \,.
\end{align}
The derivatives $p_k''(\xi_n)$ and $p_k'(\xi_n)$ are determined by~\cite{Baumann:2019eav}
\begin{align}
p_k'(\zeta_n) =&  \left\{
             \begin{array}{lr}
             \frac{q_k/q_n}{\zeta_n - \zeta_k} & \quad n \neq k   \\
            -\sum_{k\neq n} p_k'(\zeta_n) & \quad n = k  
             \end{array}
\right. \, , \\
p_k''(\zeta_n) =&  \left\{ 
             \begin{array}{lr}
            2 p_k'(\xi_n)p_n'(\xi_n) -\frac{2p_k'(\xi_n)}{\xi_n -\xi_k}& \quad n \neq k   \\
            -\sum_{k\neq n} p_k''(\zeta_n) & \quad n = k  
             \end{array}
\right. \, .
\end{align}
At any given order $N$, the algebraic system is solved in terms of an initial guess for $\omega$. The exponential rate of convergence with growing $N$ depends on the location of the closest poles in the complex plane, cf.~e.g.~\cite{Baumann:2019eav,Garcia-Saenz:2022wsl}. 
\\

%======================================
\section{Convergence of the spectral approximation}
\label{app:spectralConvergence}
%======================================

 When solving Eq.~\eqref{eq:masterEq_monopole} with spectral methods on an analytic (approximation of the) background, there are two interconnected sources or errors that determine convergence: (i) the error due to the finite order of the analytic approximation of the background (in our case a continued-fraction expansion, see App.~\ref{app:continuedFractionApprox}) and (ii) the error due to the finite number of Chebyshev nodes (see App.~\ref{app:spectral}).
 
 For the Schwarzschild background, the background spacetime, of course, has a closed analytic form and there is no error due to (i). Regarding (ii), we observe fast exponential convergence with the number of Chebyshev nodes, cf. left-hand panel in Fig.~\ref{fig:convergence}.
 
 For the non-Schwarzschild background, both sources of error are relevant, see Fig.~\ref{fig:spectral_nonSchw}: apparently the background approximation is the dominant source of error. Moreover, we observe very slow (if at all) convergence with growing number of Chebyshev nodes, cf. right-hand panel in Fig.~\ref{fig:convergence}. We suspect that this occurs due to fiducial poles in the complex plane which are introduced by the finite order of the continued-fraction approximation of the non-Schwarzschild background.
 
%\newpage
\bibliography{bibliography}% Produces the bibliography via BibTeX.

\end{document}